\documentclass[twocolumn, trackchanges]{aastex61}

\usepackage[latin1]{inputenc}
\usepackage{amsmath}
\usepackage{wasysym}
\usepackage{amsfonts}
\usepackage{amssymb}
\usepackage{graphicx}
\usepackage{listings}
\usepackage{natbib}
\newcommand\T{\rule{0pt}{2.6ex}}			%Top strut, small space above text in table
\newcommand\B{\rule[-1.2ex]{0pt}{0pt}}		%Bottom strut, small space below text in table

\shorttitle{The slowly varying corona: I}
\shortauthors{Schonfeld et al.}

\begin{document}

\title{The slowly varying corona: I --- Daily differential emission measure distributions derived from EVE spectra}

\author{S. J. Schonfeld}
\affiliation{New Mexico State University, Department of Astronomy\\ P.O. Box 30001, MSC 4500\\ Las Cruces, NM 88003-8001}

\author{S. M. White}
\affiliation{Air Force Research Laboratory, Space Vehicles Directorate\\ 3550 Aberdeen Ave SE\\ Kirtland AFB, Albuquerque, NM 87117}

\author{R. A. Hock-Mysliwiec}
\affiliation{Air Force Research Laboratory, Space Vehicles Directorate\\ 3550 Aberdeen Ave SE\\ Kirtland AFB, Albuquerque, NM 87117}

\author{R. T. J. McAteer}
\affiliation{New Mexico State University, Department of Astronomy\\ P.O. Box 30001, MSC 4500\\ Las Cruces, NM 88003-8001}

\correspondingauthor{S. J. Schonfeld}
\email{schonfsj@gmail.com}

\begin{abstract}
Daily differential emission measure (DEM) distributions of the solar corona are derived from spectra obtained by the Extreme-ultraviolet Variability Experiment (EVE) over a 4-year period starting in 2010 near solar minimum and continuing through the maximum of solar cycle 24. The DEMs are calculated using six strong emission features dominated by Fe lines of charge states \ion{}{8}, \ion{}{9}, \ion{}{11}, \ion{}{12}, \ion{}{14}, and \ion{}{16} that sample the non-flaring coronal temperature range 0.3--5 MK. A proxy for the non-\ion{Fe}{18} emission in the wavelength band around the 93.9 \AA\ line is demonstrated. There is little variability in the cool component of the corona (T $<$ 1.3 MK) over the four years, suggesting that the quiet-Sun corona does not respond strongly to the solar cycle, whereas the hotter component (T $>$ 2.0 MK) varies by more than an order of magnitude. A discontinuity in the behavior of coronal diagnostics in 2011 February--March, around the time of the first X-class flare of cycle 24, suggests fundamentally different behavior in the corona under solar minimum and maximum conditions. This global state transition occurs over a period of several months. The DEMs are used to estimate the thermal energy of the visible solar corona (of order $10^{31}$ erg), its radiative energy loss rate (2.5--8 $\times 10^{27}$ erg s$^{-1}$), and the corresponding energy turnover timescale (about an hour). The uncertainties associated with the DEMs and these derived values are mostly due to the coronal Fe abundance and density and the CHIANTI atomic line database.

\end{abstract}


\section{Introduction}
\label{sec:introduction}
The solar corona (the outer layer of the Sun's atmosphere) plays an important role in solar activity and the Sun's impact on the Earth's atmosphere. The high (million degree K) temperatures found in the corona result from a still unidentified \citep[but likely magnetic-field dominated, e.g.,][]{Zirker1993, Walsh2003, Klimchuk2006} heating mechanism that must be a fundamental process, since it is known to occur across a wide range of stellar types. The distribution of energy with temperature in the corona presumably reflects the nature of this mechanism and the way that energy is redistributed through the corona from the locations where heat is deposited.

In principle it is simple to determine the distribution of coronal plasma with temperature (known as the ``differential emission measure'', or DEM) by inverting a set of temperature-sensitive extreme-ultraviolet (EUV) observations. In practice however, observational noise, finite observations, and incomplete knowledge of the relevant atomic physics make this an ill-posed problem complicated by the computational details of the solution algorithm. These difficulties have been well understood for decades \citep{Craig1976} and considerable effort is still being made to validate these DEM analyses \citep{Guennou2012a, Guennou2012b,Testa2012a, Aschwanden2015}. Furthermore, recent years have seen the advent of impressive new DEM calculation techniques \citep{Hannah2012, Plowman2013b, Cheung2015}. These have been validated for a wide range of coronal conditions and run quickly on modern computers, allowing DEM studies of larger spatial and temporal domains than ever before.

The Solar Dynamics Observatory (SDO), launched in 2010 \citep{Pesnell2011}, has led to greatly improved understanding of the solar corona including determination of coronal DEMs with both images at several EUV passbands from the Atmospheric Imaging Assembly \citep[AIA;][]{Lemen2012} and spectral irradiance measurements from the EUV Variability Experiment \citep[EVE;][]{Woods2012}. This has been accomplished for studies of solar flares \citep[e.g.,][]{Hock2012a, Fletcher2013, Kennedy2013, Warren2013a, Caspi2014, Warren2014a, Zhu2016}, active regions \citep[e.g.,][]{Warren2012, Aschwanden2013a, DelZanna2013b, Petralia2014}, coronal loops \citep[e.g.,][]{Aschwanden2011, Warren2011, DelZanna2011b}, the full Sun \citep[e.g.,][]{Nuevo2015, Schonfeld2015}, and even the entire corona over a complete Carrington rotation \citep{Vasquez2016}. Major advances provided by SDO include consistent, high temporal resolution, long-term, full-Sun observations.

In this paper we present a study of the long-term coronal DEM variability leveraging these uniform data sets over a significant fraction of the solar cycle. Considering the corona in such a holistic sense provides perspectives lost in narrowly focused active region studies. EVE spectra are particularly well suited to this task because extra effort has been made to provide in-flight calibration thanks to sounding rocket under-flights with an identical instrument \citep{Hock2010}. Additionally, the ability to identify individual emission lines in EVE spectra allows for the selection of diagnostics representing a wide range of coronal conditions. We present an analysis of the variation of the corona over a significant fraction of the solar cycle through calculation of daily full-Sun integrated DEMs utilizing the complete EVE MEGS-A data set. We describe the instrument and data set in \S \ref{sec:EVE}. A description of the DEM calculation as well as relevant underlying assumptions is given in \S \ref{sec:DEM} and the DEM validation is discussed in \S \ref{sec:validation}. Implications of the results on the coronal energy content and its evolution are discussed in \S \ref{sec:evolution}. We conclude and discuss future uses of these results in \S \ref{sec:conclusion}. We also discuss an analysis of the solar spectrum near the \ion{Fe}{18} 94 \AA\ line in Appendix \ref{sec:Fe18}.

\section{EVE MEGS-A Coronal Spectra}
\label{sec:EVE}

The EUV Variability Experiment (EVE) includes a suite of instruments designed to observe the solar EUV irradiance from 1--1050 \AA\ with high cadence, spectral resolution, and accuracy. Within this suite, the Multiple EUV Grating Spectrographs (MEGS)-A grazing-incidence spectrograph observed the solar irradiance over the wavelength range 50--370 \AA\ with better than 1 \AA\ resolution and greater than 25\% irradiance accuracy \citep{Woods2012}. MEGS-A operated nearly continuously from 2010 April 30 until 2014 May 26 when it suffered a CCD failure \citep{Pesnell2014EVE}. There were four CCD bake-out procedures during this period when no data were collected\footnote{Bakeouts occured in the periods 2010 June 16--18, 2010 September 23--27, 2012 March 12--13, and 2012 March 19--20.}.

\begin{figure*}[t] %fig:spectrum
	\centering
	\includegraphics*[trim=0cm 0cm 0cm 0cm, scale=1.0]{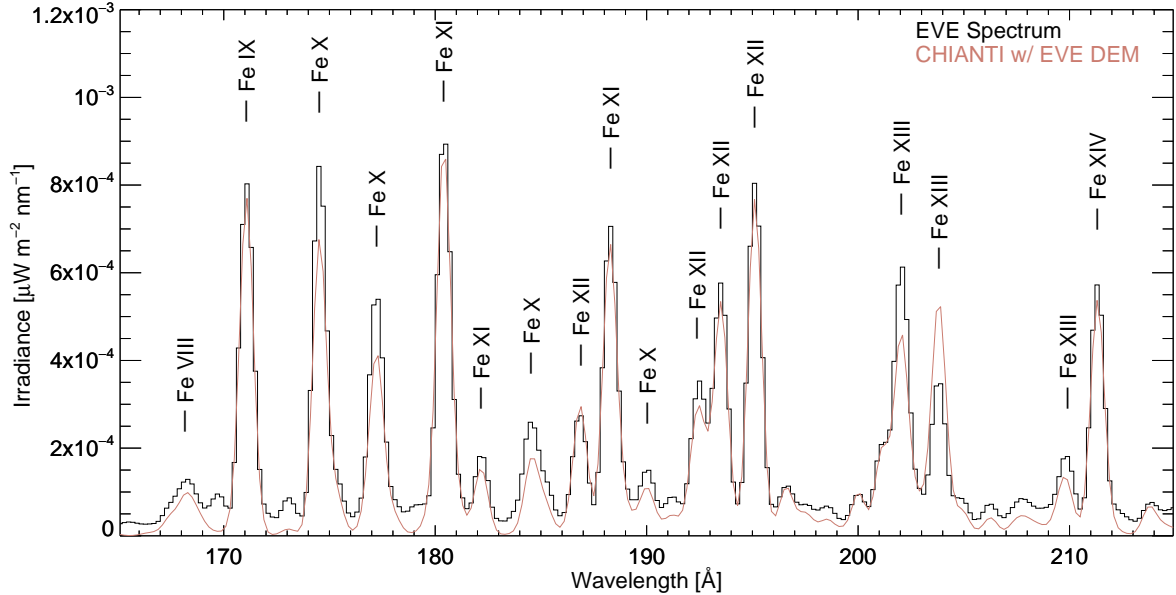}
	\caption{A portion of the observed EVE spectrum on 2011 November 6 (black histogram) along with the CHIANTI model spectrum (red line) calculated using the DEM computed as described in \S \ref{sec:DEM}. The CHIANTI lines have been convolved with a 0.75 \AA\ full-width at half-maximum Gaussian to generate the model spectrum. This rich region of the spectrum contains many strong emission lines from \ion{Fe}{8}\ --\ion{}{14} that originate in the corona. Note that the synthetic CHIANTI spectrum is believed to lack a large number of weak unresolved lines that appear in the EVE spectra as an offset, which partly explains why some of the lines appear stronger in the EVE spectrum than in the CHIANTI spectrum.}
	\label{fig:spectrum}
\end{figure*}                                                                                                                                                                             

For this study we use the MEGS-A spectra collected every day between 19:00 and 20:59 UT\footnote{This interval is chosen to match the timing of the daily F$_{10.7}$ measurement at 20 UT.} to compute a representative ``daily'' spectrum. In practice we use the median in each 0.2 \AA\ MEGS-A wavelength bin over the two hour period (comprising 720 spectra taken at 10 second intervals) to create a median spectrum. Use of the median minimizes the effects of short-timescale variability, including flares, during the observation window. Long-duration flares will still perturb the median values that we use. We make no attempt to remove such events, or global coronal changes on hour-long timescales, from the spectra because we consider them important aspects of the long-term coronal evolution. All of our analysis is performed using these daily median spectra.

As an example of typical daily median MEGS-A data, Figure \ref{fig:spectrum} shows the EVE spectrum in the wavelength range 165--215 \AA , which contains a large number of strong coronal emission lines. Important to note here is that the resolution of the EVE spectra does not resolve the intrinsic width of the coronal lines. The typical full-width at half-maximum of lines we measure in the MEGS-A spectra is $\sim 0.75$ \AA\ \citep[although the instrument line width was found to be $\sim 0.85$ \AA\ by][]{Hock2012}, while the actual intrinsic line widths are of order 0.1 \AA\ \citep{Feldman1974}. Nonetheless, strong coronal lines such as those labeled in Figure \ref{fig:spectrum} are usually clearly visible in the spectra. We use Version 8.0.2 of the CHIANTI atomic line database \citep{Dere1997,DelZanna2015a} for line identification.

\begin{figure*}[ht] %fig:lines
	\centering
	\includegraphics*[trim=0cm 0cm 0cm 0cm, scale=1]{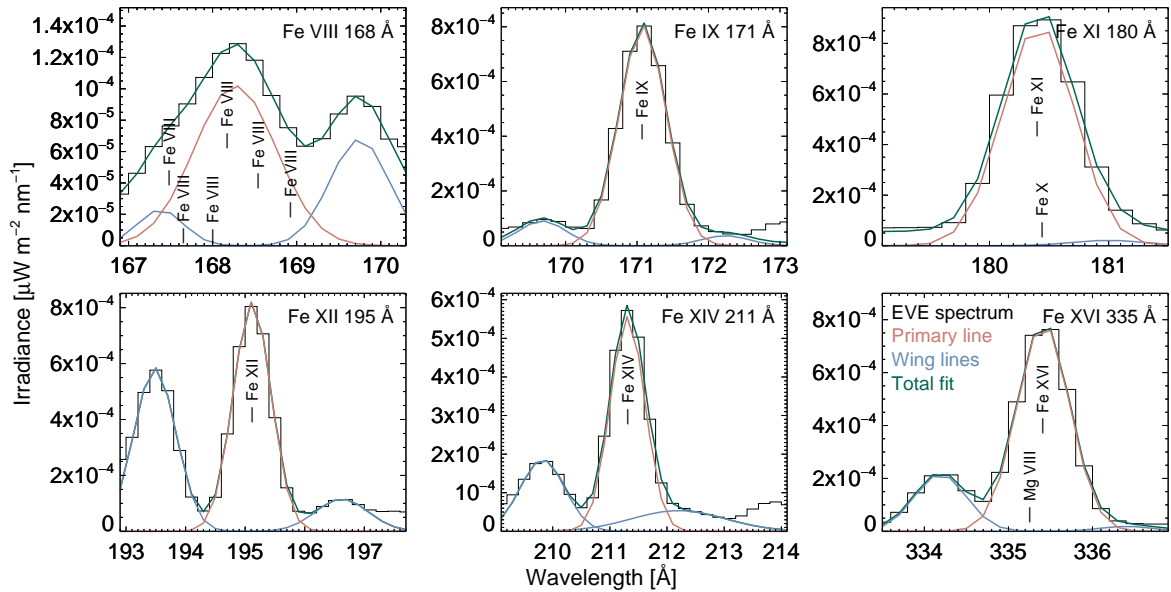}
	\caption{The EVE spectrum and associated Gaussian fits on 2011 November 6 for the bold lines in table \ref{table:lines} used in the DEM calculations. The black histogram is the observed median EVE spectrum. The red and blue lines are the Gaussian fits for the primary line and the wing features respectively. The green line is the total spectral fit including the three Gaussians and a constant background. The associated contributing lines listed in table \ref{table:lines} are also indicated at the proper wavelength and relative strength.}
	\label{fig:lines}
\end{figure*}                                                                                                                                                                             

In order to generate accurate and consistent DEMs from MEGS-A spectra we identify a list of suitable candidate lines, i.e. strong features in the spectra believed to be dominated by individual spectral lines. For the purpose of deriving a complete census of coronal emission as a function of temperature, we require lines covering the broadest possible temperature range above about 0.3 MK. In view of the lingering debate surrounding coronal abundances in relation to the first ionization potential (FIP) effect \citep{Feldman1992,White2000a,Asplund2009}, we chose to restrict our analysis to Fe emission lines (most of the strong lines in the MEGS-A spectra) in order to minimize the number of elemental abundances required for our calculations. Further details regarding the effects of elemental abundance are discussed in \S  \ref{sec:abundance}.

MEGS-A spectra contain strong lines for all Fe charge states in the range \ion{}{8}\ --\ion{}{16}, as well as \ion{}{18}. The peak temperatures of the responses of this set of Fe charge states cover the range 0.6--7.1 MK, i.e., they sample the bulk of the non-flaring corona. The list of strong, relatively isolated MEGS-A emission features dominated by the emission from a single stage of Fe, with their primary contributing transitions, peak temperature, and relative strength determined from CHIANTI, is given in Table \ref{table:lines}. For each line we also identify features blended within the full-width at half-maximum of the target transition. All of the EVE features are almost pure Fe emission, with the exception of \ion{Fe}{16} 335 \AA\, which has a significant \ion{Mg}{8} line 0.15 \AA\ blue-ward of the primary line.

To extract the flux of each primary line we fit the emission features in the median daily spectra with three Gaussian functions (four in the case of \ion{Fe}{18}), one at the primary wavelength and one each in the red and blue wings to account for the flux from neighboring emission features. The wavelength, width, and strength of each Gaussian component is allowed to vary in the fitting process, although the allowed wavelength range is constrained in some cases (notably for \ion{Fe}{8} 168 \AA) and the width of the wing features is constrained to the width of the primary feature when the wings lack identifiable peaks. A characteristic set of line fits are shown in Figure \ref{fig:lines}. For each line the observed flux is taken as the integrated flux in the primary Gaussian, with the exception of the \ion{Fe}{8} 168 \AA\ feature, which is actually a complex of six \ion{Fe}{8} lines. For \ion{Fe}{8} 168 \AA\ the flux in the blue-wing Gaussian is added to the flux in the primary line since the wing is also dominated by \ion{Fe}{8} emission. The uncertainties in the fitted fluxes are determined from the uncertainties in the line amplitude and width found during the fitting procedure.

The line fits also include a constant background component. This is included to account for weak lines that aren't included in the CHIANTI database but that must be present in the spectrum to account for the offsets in the minimum flux level observed in MEGS-A spectra. The most obvious example of this is the \ion{Fe}{18} 94 \AA\ line explored in Appendix \ref{sec:Fe18}. True continuum emission in this region of the EUV spectrum is negligible for the non-flaring Sun, accounting for well less than 1\% of the emission in any individual line.

\section{DEM Assumptions and Calculation}
\label{sec:DEM}

In this section we describe the choices made in deriving daily full-Sun DEMs from EVE MEGS-A data. The DEM with units of $\text{cm}^{-3}\ \text{K}^{-1}$ is defined as:

\begin{equation}
\label{eqn:DEM}
\text{DEM(T)} = \int\displaylimits_{\text{V}}\frac{d}{d\text{T}}\left(n_{\text{e}}n_{\text{H}}\right) d\text{V}
\end{equation}
where $n_{\text{e}}$ and $n_{\text{H}}$ are the electron and proton number densities, respectively, T is the coronal electron temperature, and the integral is over the visible coronal volume V. EVE measures the irradiance $\left(\text{W}\ \text{m}^{-2}\ \text{nm}^{-1}\right)$ while CHIANTI performs calculations natively using radiance $\left(\text{erg}\ \text{cm}^{-2}\ \text{s}^{-1}\ \text{sr}^{-1}\right)$. EVE's field of view is several degrees wide, and it has no spatial resolution, hence it does not measure the actual solid angle of the solar emission. We choose to provide radiances to CHIANTI by dividing the observed EVE irradiances by the solid angle occupied by the area of the solar disk at 1 AU, $6.78\times 10^{-5}$ sr, which conveniently gives us quantities comparable to spatially resolved DEM analyses. CHIANTI (see \S \ref{sec:calculation}) then returns the averaged column DEM $\left(\text{cm}^{-5}\ \text{K}^{-1}\right)$ which is just the total volume emission measure $\left(\text{cm}^{-3}\ \text{K}^{-1}\right)$ of the Sun divided by the area of the solar disk. Multiplying this column DEM by the area of the solar disk directly cancels the arbitrary division by the solid angle of the solar disk described above and yields the volume-integrated DEM of the solar corona.

\subsection{Choice of lines for DEM fitting}
\label{sec:lines_used}

Selection of suitable lines for DEM analysis is critical because the detailed atomic characteristics associated with the chosen emission lines must be fully characterized in order to properly calculate the DEM. We therefore use only those emission lines that are most well characterized for our analysis.

Since the lines of a given charge state in Table \ref{table:lines} have essentially identical contribution functions (emission as a function of temperature), the use of more than one line per charge state does not add more information in the DEM analysis. Use of multiple lines from the same charge state can over-weight the corresponding temperature range compared to states with a single line available, and in addition may hinder the fitting procedure if the lines have different density dependencies. Therefore, we choose to use only a single line for each charge state in the DEM fitting. On the basis of line strength we use \ion{Fe}{8} 168 \AA\ rather than \ion{Fe}{8} 131 \AA\ and \ion{Fe}{11} 180.4 \AA\ rather than \ion{Fe}{11} 188.2 \AA\ for the DEM fits.

\begin{figure*}[tp] %fig:density_dependence
	\centering
	\includegraphics*[trim=0cm 0cm 0cm 0cm, scale=1.0]{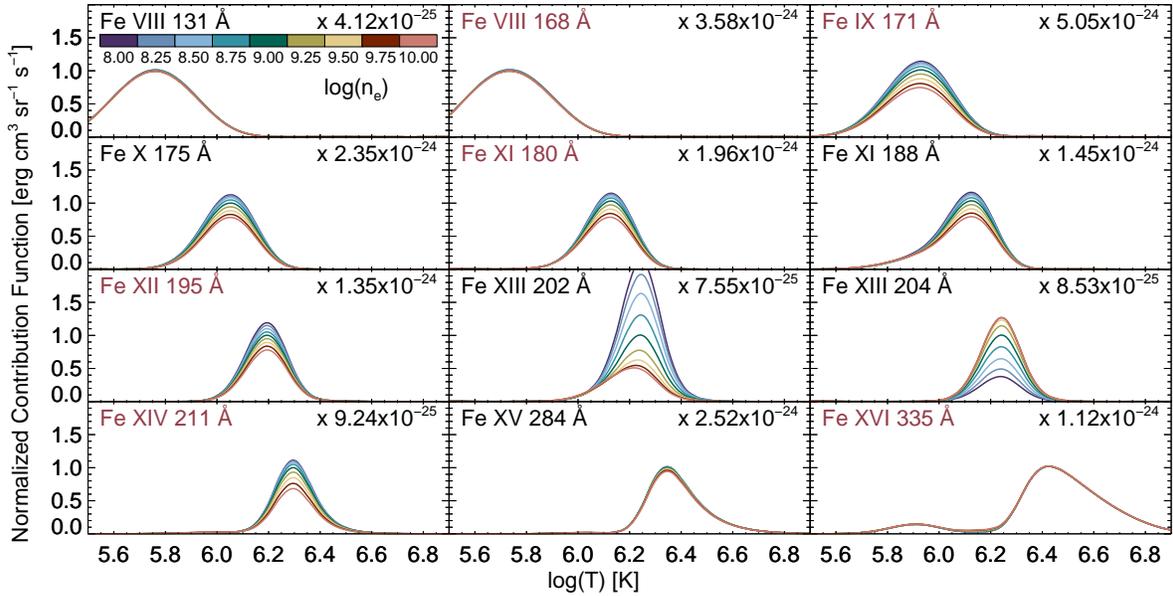}
	\caption{The contribution functions (intensity per emission measure as a function of temperature) with units of $\text{erg}\ \text{cm}^{3}\ \text{s}^{-1}\ \text{sr}^{-1}$ of the features in Table \ref{table:lines} (excluding Fe XVIII). Curves are plotted for nine different values of coronal density in the range 10$^8$--10$^{10}$ cm$^{-3}$ and normalized to the peak emission at 10$^{9}$ cm$^{-3}$. Lines labeled in red are used to compute the DEMs.}
	\label{fig:density_dependence}
\end{figure*}                                                                                                                                                                             

The density dependence of a given line is also an important consideration in the line choice. EUV emission lines are all collisionally excited and their emission properties depend on density \citep{Mason1994}. A density must therefore be specified to determine the temperature response of lines used in DEM calculations. Ideally, the lines used for DEM fitting will all have similar density dependence, but in practice the limited number of lines available to choose from means that this is not generally possible. Not surprisingly, using a combination of lines with very different density dependencies typically produce poor DEM solutions. Figure \ref{fig:density_dependence} shows the density variation of the lines in Table \ref{table:lines} over the plausible coronal range 10$^8$--10$^{10}$ cm$^{-3}$. The \ion{Fe}{8}, \ion{}{15}, and \ion{}{16} (and \ion{}{18}, not shown) lines all have essentially no density dependence over this range, while \ion{Fe}{9}, \ion{}{10}, \ion{}{11}, \ion{}{12}, and \ion{}{14} all show some variation with density but have very similar behavior, with increased emission at lower density. Both \ion{Fe}{13} lines, however, show a dramatic change in emission with density, and in practice we find that inclusion of these \ion{Fe}{13} lines produces very poor DEM solutions. For this reason we exclude them from the DEM calculation, but they  provide a useful density diagnostic that we discuss further in \S \ref{sec:density} and \ref{sec:DEM_validation}.

The \ion{Fe}{10} 175 \AA\ and \ion{Fe}{15} 284 \AA\ lines are both strong, relatively isolated emission features that could be expected to be valuable in constraining the DEM. However, the inclusion of either of these lines leads to dramatic fluctuations in the DEM calculations including the appearance of sharp reductions of emission measure at 1 MK and 2.5 MK (log(T) = 6.0 and 6.4) and generally poor reproduction of the input data. We regard sharp features in the DEM as unphysical because the radiative loss functions (discussed further in \S \ref{sec:evolution}) are smooth functions of temperature and we have no evidence that coronal heating favors narrow temperature ranges. Additionally, as discussed in \S \ref{sec:DEM_validation}, the \ion{Fe}{10} 175 \AA\ line shows evidence of systematic errors in its representation in CHIANTI. We, therefore, exclude the \ion{Fe}{10} 175 \AA\ and \ion{Fe}{15} 284 \AA\ lines from the DEM fitting procedure.

The \ion{Fe}{18} line at 94 \AA\ can provide an important constraint on the DEM at high temperatures (up to 10 MK), but we find \citep[in agreement with e.g.,][]{Aschwanden2011, Reale2011, Testa2012, Aschwanden2013a} that CHIANTI presently does not represent the relevant region of the EUV spectrum sufficiently well to rely on the \ion{Fe}{18} line. We discuss this issue specifically for EVE spectra in more detail in Appendix \ref{sec:Fe18}, including the discovery of a proxy for the non-\ion{Fe}{18} component in this wavelength range.

The DEM fits are derived here using the \ion{Fe}{8}, \ion{}{9}, \ion{}{11}, \ion{}{12}, \ion{}{14}, and \ion{}{16} lines marked in bold in Table \ref{table:lines}. These lines span the non-flaring coronal temperature range from 0.3 to 5 MK with sufficient sensitivity over the whole range to suitably represent the solar coronal DEM.

\begin{figure}[t] %fig:G(T)
	\centering
	\includegraphics*[trim=0cm 0cm 0cm 0cm, scale=0.9]{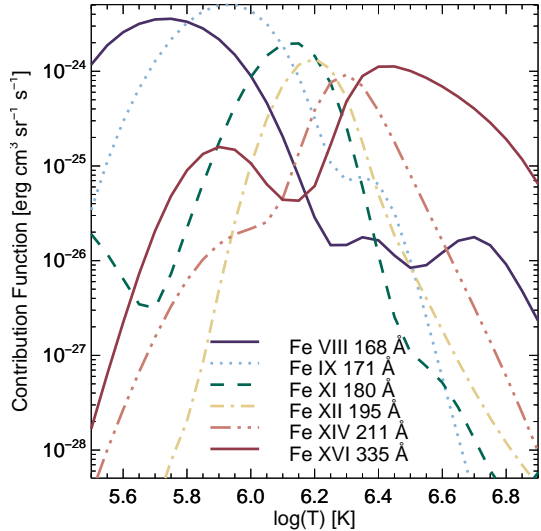}
	\caption{The contribution functions of the lines in Table \ref{table:lines}. This shows the temperature sensitivity of the individual emission lines and the combined sensitivity of the DEM. Notice that even though these lines (except for Fe XVI 335 \AA) can be considered isothermal they have significant emission over a range of temperatures.}
	\label{fig:G(T)}
\end{figure}                                                                                                                                                                             

The corresponding contribution functions (intensity per emission measure as functions of temperature) used to calculate the DEMs are shown in Figure \ref{fig:G(T)}. These are generated by summing the contribution from each CHIANTI emission line (of which the strongest contributors are listed in Table \ref{table:lines}) within the full-width at half-maximum centered on the wavelength of each individual EVE feature, assuming coronal elemental abundances \citep{Feldman1992}.

\subsection{Coronal density}
\label{sec:density}
Since the DEM itself represents an integral of the square of plasma density along lines of sight through the solar atmosphere, it does not contain specific information about the density at any point. However, electronic excitation and a fraction of the de-excitation in the corona is caused by electron-ion collisions whose rate is mediated by the electron density \citep{Gaetz1983}. Therefore, the rates at which individual excitation states within an ion are populated and de-populated are functions of the electron density. Depending on the details of the excitation and de-excitation pathways for each individual transition, increased density can lead to decreased (through collisional quenching) or increased (through collisional excitation) emission. Different transitions of the same charge state can have different dependencies on density (e.g., the \ion{Fe}{13} lines in Figure \ref{fig:density_dependence}), a property that is exploited for coronal density diagnostics \citep{Tripathi2008, Warren2009, Young2009}. Accordingly, the choice of density used to determine the responses of different EVE features can produce quantitative changes in the DEM results.

We, therefore, must choose a density to use when calculating the temperature responses of the lines supplied for DEM fitting. In active regions, non-flaring densities can be as high as 3--10 $\times 10^{10}\ \text{cm}^{-3}$ \citep{Tripathi2008, Young2009}, while in the diffuse quiet corona outside active regions values as low as 6--25 $\times 10^7\ \text{cm}^{-3}$ \citep{Doschek1997, Warren2009} may be appropriate. The strongest density-sensitive lines in the EVE MEGS-A range are the \ion{Fe}{13} 202 \AA\ and 204 \AA\ lines, as shown in Fig. \ref{fig:density_dependence}. We use CHIANTI to determine the density corresponding to the fluxes in these two lines in typical EVE daily spectra and they suggest a density of $10^{8.5-9}\ \text{cm}^{-3}$. Noting the fact that EUV emission is proportional to density squared and therefore will always be biased towards higher densities, we adopt $10^{9.0}\ \text{cm}^{-3}$ as the density for our calculations. In order to account for the effect of this density choice on the results we also perform the DEM calculations using $10^{8.5}$ and $10^{9.5}\ \text{cm}^{-3}$ and use the resulting variation in the DEMs as a measure of the uncertainty in our final DEMs.

It must be noted that a single electron density is certainly not appropriate to describe the global corona. This is because, for example, high temperature lines will preferentially originate from active regions where we expect the density to be higher than in the quiet Sun where lower temperature lines dominate. However, without a formal quantitative basis on which to assign different densities to different charge states, we choose to use a common density for all the lines employed in the DEM calculations. It is also likely that the average coronal density will change with the solar activity level. With only the single pair of density sensitive \ion{Fe}{13} lines however, we do not have enough independent constraints on the temporal density evolution to vary the assumed coronal density with time in the DEM calculations.

\subsection{Abundances}
\label{sec:abundance}
The energetics of the solar corona are dominated by the most populous elements, hydrogen (which is a single proton at coronal temperatures) and helium (an alpha particle). Thus, the emission measure of interest is the total emission measure, dominated by hydrogen, helium, and the electrons they donate to the plasma. However, H and He do not produce lines in the EUV that are useful for determining coronal DEMs whereas Fe, as discussed above, has a large number of suitable lines. Therefore, by using Fe emission lines we actually solve for the DEM of Fe and convert it to a total DEM, correcting for the abundance by multiplying by N$_{\text{H}}$/N$_{\text{Fe}}$. By using only emission lines from various stages of iron in our DEM calculations we, to first order, simplify the influence of elemental abundances on the DEM down to a single value, the Fe abundance. This neglects the influence of secondary emission from other elements (such as the Mg VIII contribution to the Fe XVI line), but as these contributions are quite small their influence is likely negligible. This means that the total DEM calculated from purely Fe emission lines scales inversely with the Fe abundance, assuming the abundance is constant throughought the solar corona. This analysis uses the standard ``coronal'' iron abundance of $\text{N}_{\text{Fe}}/\text{N}_{\text{H}} = 1.26\times10^{-4}$, four times that of the photosphere \citep{Feldman1992}, which \cite{Schonfeld2015} demonstrated to be suitable for full disk coronal analysis with emission dominated by active regions.

\subsection{DEM calculation}
\label{sec:calculation}

\begin{figure*}[t] %fig:DEM
	\centering
	\includegraphics*[trim=0.9cm 0cm 0.5cm 0cm, width=\textwidth]{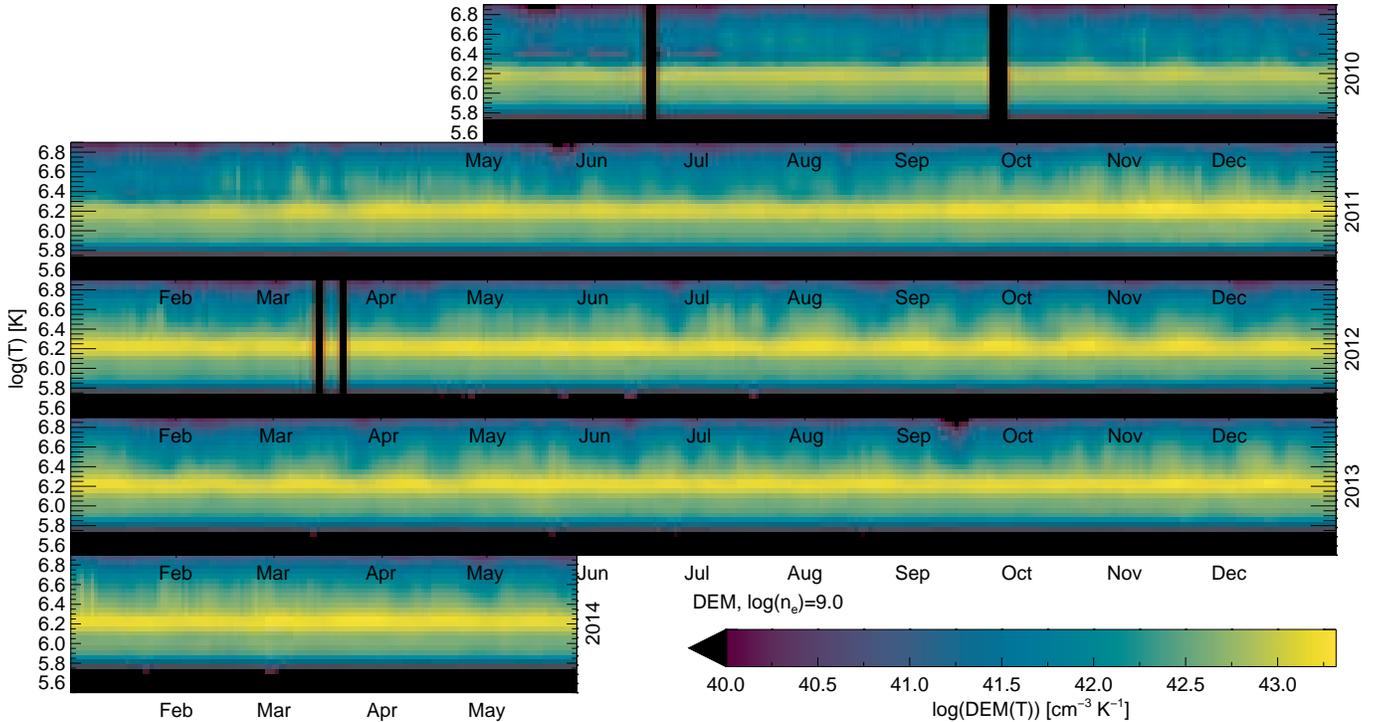}
	\caption{The DEM time series calculated for the complete EVE data set. The vertical black bands in 2010 and 2012 are the result of data gaps explained in \S \ref{sec:EVE}. The plot shows an increase in high temperature plasma (log(T) $\ge$ 6.3) associated with solar maximum (2011--2014). Rotational modulation is evident throughout the observations, most clearly in the period 2012 July -- 2013 April.}
	\label{fig:DEM}
\end{figure*}                                                                                                                                                                             

We use the line fluxes and uncertainties extracted from the median MEGS-A spectra to generate daily full-Sun-integrated coronal DEMs. These DEMs are derived using Version 8.0.2 of the CHIANTI database. We use the regularized-inversion DEM solution method from \cite{Hannah2012} as implemented in CHIANTI, restricting the solutions to the temperature range $5.5 \leq$ log\text(T) $\leq 6.9$ with bins of log(T) = 0.05. We choose to enforce positivity in the DEM solutions to prevent nonphysical negative emission measures, but in practice we find that the solutions obtained using the six chosen EVE features are uniformly positive without this constraint (which is not the case when other lines in Table \ref{table:lines} are included). The full four-year DEM time series resulting from our analysis is shown in Figure \ref{fig:DEM}.

\section{DEM Validation}
\label{sec:validation}

The DEMs show a clear increase in coronal activity from near solar minimum in $2010$ to solar maximum in $2011$--$2014$ including a slight increase in the peak temperature. During solar maximum there are times when a pronounced and consistent rotational modulation signal is present (particularly $2012$ July--$2013$ April), indicating a relatively stable corona with strong active regions in fixed longitude ranges regularly rotating on and off the visible disk. However, there are also times when the solar activity loses that regularity and the rotational signal becomes obscured, such as during $2013$ June--November.

\subsection{Uncertainty estimates}
\label{sec:uncertainty}

\begin{figure*}[!ht] %fig:compare
	\centering
	\includegraphics*[trim=0cm 0cm 0cm 0cm, scale=1]{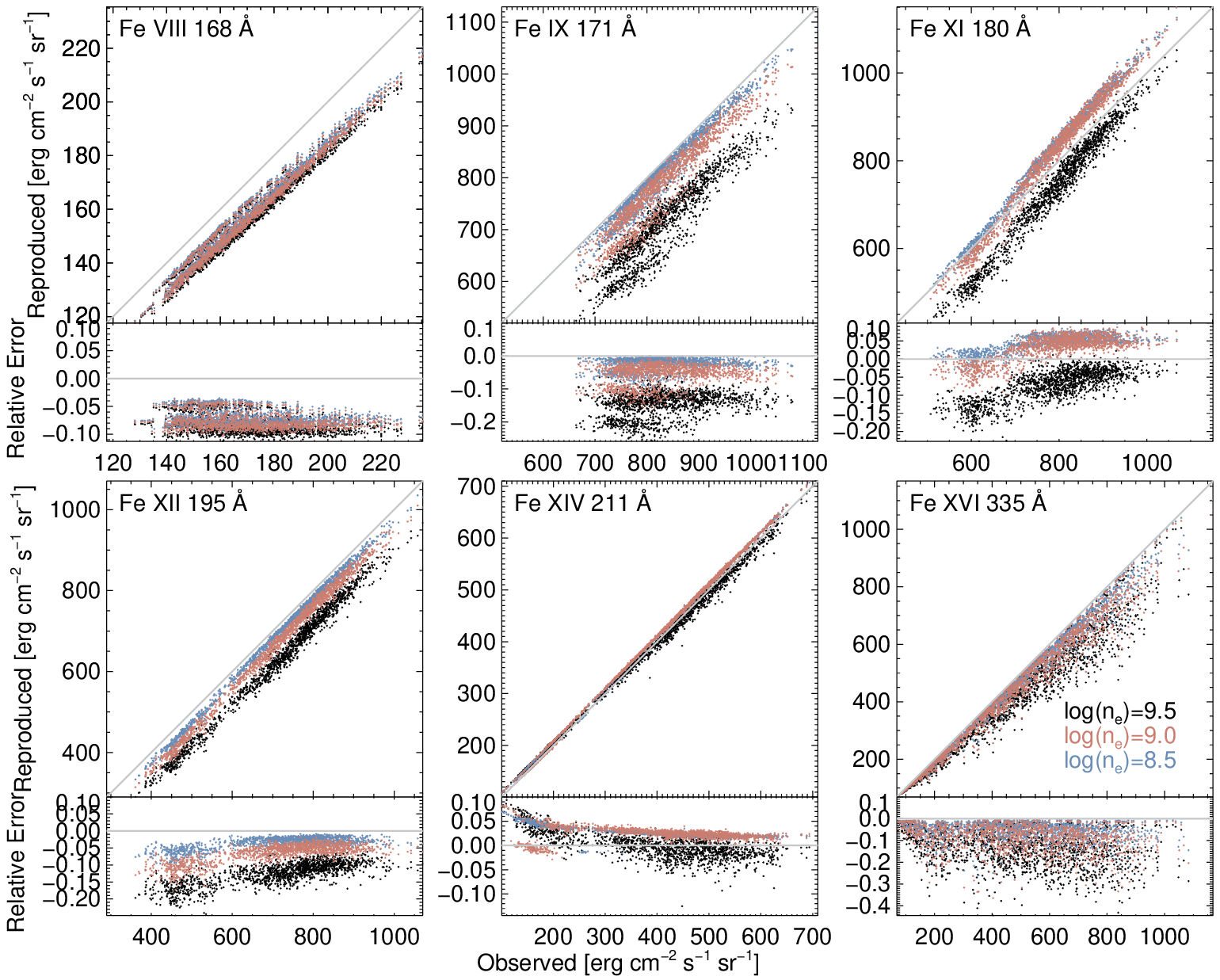}
	\caption{The fluxes reproduced by the calculated DEMs vs. the observed flux for the lines used to compute the DEMs. The diagonal gray lines indicate where the reproductions equal the observations. A clear trend of reduced reproduced line flux with increased density is seen across all lines. At $10^{9.0}\ \text{cm}^{-3}$, all of the lines are reproduced to better than 20\%.}
	\label{fig:compare}
\end{figure*}                                                                                                                                                                             

The DEM fitting procedure involves a $\chi^2$-minimization in which the emission measure in each temperature bin is adjusted such that convolving the DEM with the temperature responses of each of the six EVE features (Figure \ref{fig:G(T)}) produces model line fluxes that match the input line fluxes to within the specified measurement uncertainties. For completeness, we use the derived DEMs to compute daily synthetic EVE spectra with an example shown in Figure \ref{fig:spectrum}. This is done by summing the contribution from each individual emission line in the spectral range using the calculated DEM. We then fit the emission lines with the same procedure as was used to fit the original EVE spectrum, but this time without the constant background component (since that was only added to account for lines not included in CHIANTI). A comparison of these derived output fluxes using the three chosen densities with the input EVE fluxes is shown in Figure \ref{fig:compare}. The residual plots show that for $10^{9.0}\ \text{cm}^{-3}$: \ion{Fe}{14} is reproduced to about 5\%; \ion{Fe}{8}, \ion{Fe}{9}, \ion{Fe}{11}, and \ion{Fe}{12} to about 10\%; and \ion{Fe}{16} to better than 20\%. The systematic and consistent values of these offsets over a wide range of solar activity levels suggests that they are dominated by inconsistencies in the atomic data used to derive the response of each line and/or fundamental precision errors in the EVE MEGS-A calibration. We conclude from these results that the overall uncertainty associated with the DEM fitting is of order 10\%. This is consistent with the uncertainties reported by the fitting procedure which for the individual log(T) = 0.05 bins with significant emission measure (i.e., bins above log(T) = 5.8) are of order 10\%.

This algorithmic uncertainty associated with the line fitting and DEM calculation ignores many subtle complications in the analysis. The following additional sources of uncertainty contribute to the final estimation of the accuracy of our DEMs. 

\begin{figure*}[t] %fig:Emission Measure
	\centering
	\includegraphics*[trim=0cm 0cm 0cm 0cm, scale=1]{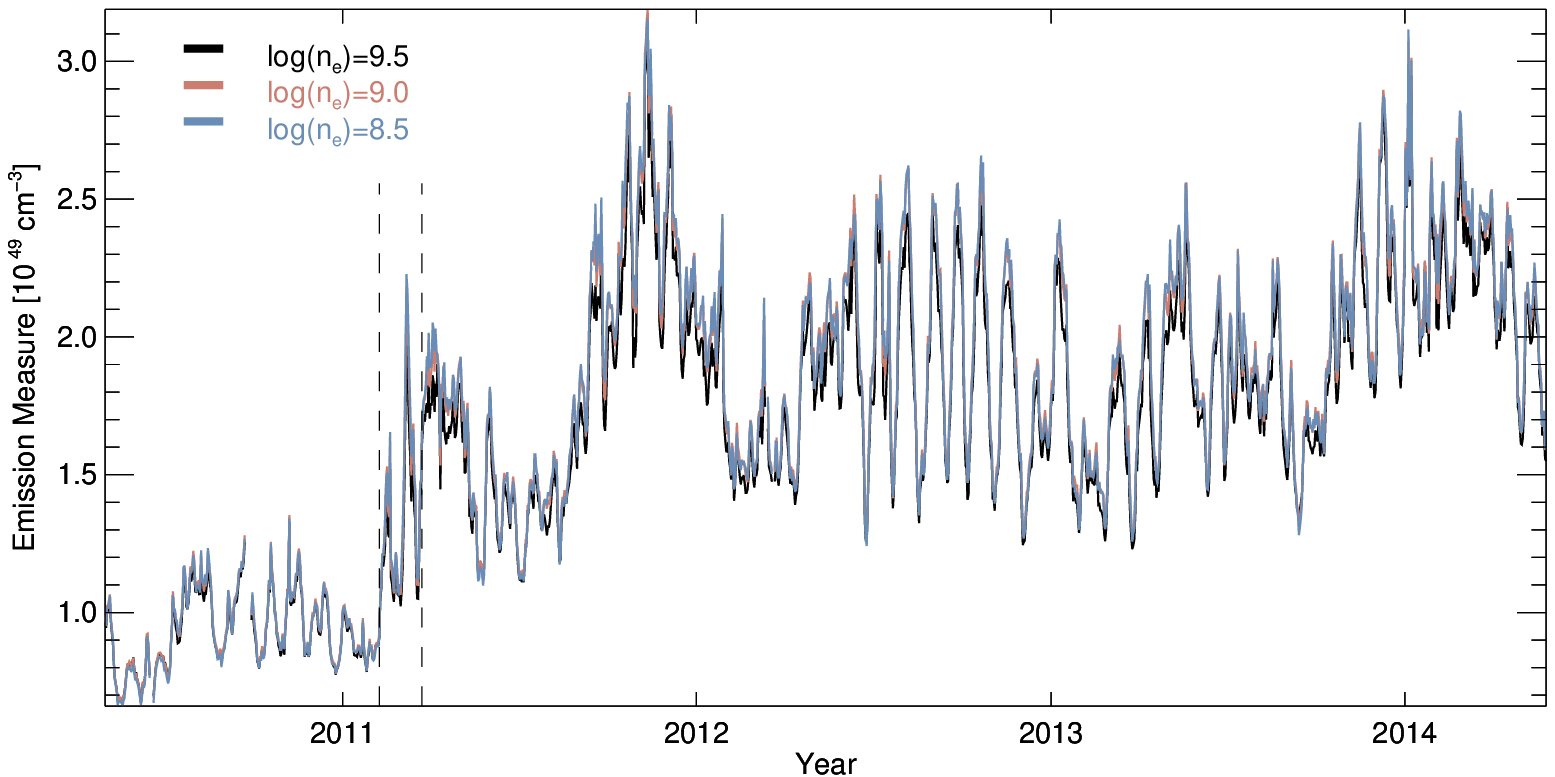}
	\caption{The total emission measure series for the complete EVE data set for each of the three tested plasma densities. The deviation of $10^{8.5}$ and $10^{9.5}$ from $10^{9.0}\ \text{cm}^{-3}$ is typically less than 5\%, with the largest deviation during periods of maximum activity. The dashed vertical lines indicate the period of transition from solar minimum to solar maximum discussed in \S \ref{sec:two-state}: they indicate the dates 2011 February 7 and 2011 March 23.}
	\label{fig:EM}
\end{figure*}

\begin{itemize}
	\item The time variability of the spectra within the two-hour window used to determine the median daily spectrum. We calculate standard deviations in each wavelength bin over the two hours for each day and find that variations at the peaks of the strong cooler lines (\ion{Fe}{9} 171 \AA , \ion{Fe}{11} 180 \AA , and \ion{Fe}{12} 195 \AA ), which should represent temporal variability, are typically about 1\%.
	
	\item The calibration of the EVE MEGS-A irradiance spectra. \cite{Hock2012} discusses the calibration of MEGS-A in detail: the responsivity (conversion of detector counts to irradiance) is estimated to have an uncertainty better than 1\% for most of the wavelength range that we use, but possibly worse in the range 150-170 \AA\ where the A1 and A2 slit responses overlap. The irradiance calibration precision is in the range of $\pm$5--7\% for the strong MEGS-A lines we consider.

	\item The determination of line fluxes by fitting Gaussians to the EVE spectra. The formal uncertainties in these fits is a few percent, depending on the line.

	\item Uncertainty due to the need to choose a density in calculating the temperature responses of each line. Figure \ref{fig:EM} shows the total emission measure (the integral of the DEM over temperature) for three different values of density for which calculations were carried out. The spread in the resulting emission measures is $\pm$5\% which we take to be the uncertainty associated with the choice of density.

	\item Uncertainties in the atomic data used by CHIANTI to derive the emissivity and temperature response of the lines used for the DEM determination. As discussed in \S \ref{sec:DEM_validation}, there are clear discrepancies between the lines used and other strong lines in the EVE spectra. Assigning a formal uncertainty for the specific lines used to obtain the DEMs is non-trivial and not addressed here.

	\item Uncertainty in the chosen abundance of Fe. As described in \S \ref{sec:abundance}, a change in this value results in a scale change in the DEMs rather than an uncertainty. It is possible that the appropriate value of the abundance may vary with solar activity levels, and we hope to address that question in a future study.
\end{itemize}

In summary, we derive an overall uncertainty in the DEMs of order 15\%, with the recognition that uncertainties in the atomic data and the assumed Fe abundance are additional factors not well represented in that number.

\subsection{DEM testing}
\label{sec:DEM_validation}

\begin{figure*}[!ht] %fig:compare_test
	\centering
	\includegraphics*[trim=0cm 0cm 0cm 0cm, scale=1]{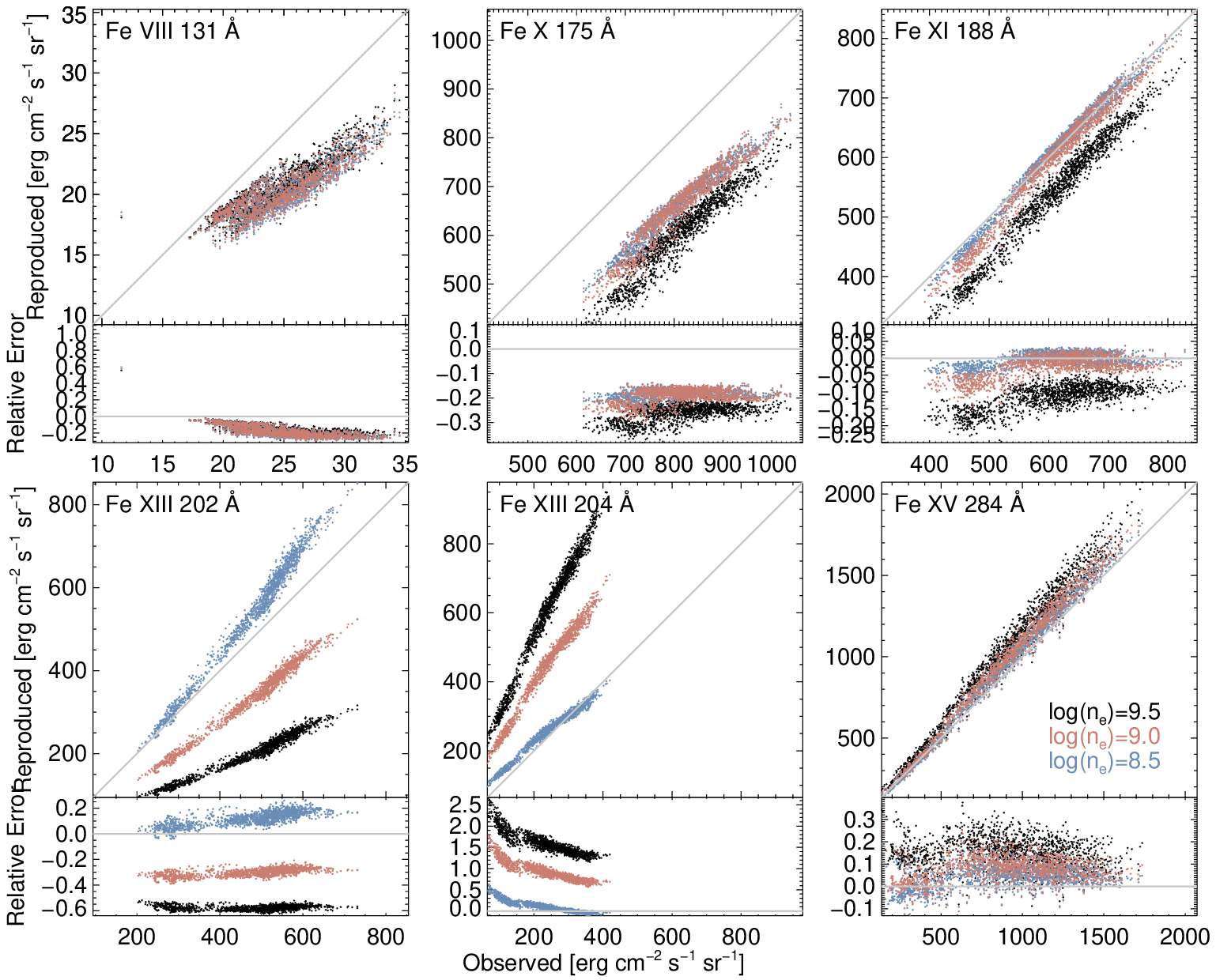}
	\caption{The same as Figure \ref{fig:compare} but for the lines in Table \ref{table:lines} not used to compute the DEMs (except \ion{Fe}{18}). \ion{Fe}{8}, \ion{Fe}{10}, and \ion{Fe}{15} all show some level of consistent deviation, with \ion{Fe}{10} being particularly striking given the high precision but low accuracy of the flux reproductions. \ion{Fe}{11} is reproduced extremely well. Both \ion{Fe}{13} lines show clear variation with density, as expected, and are consistent with the tested density range.}
	\label{fig:compare_test}
\end{figure*}                                                                                                                                                                             

As a test of the DEM accuracy Figure \ref{fig:compare_test} gives 	comparisons similar to those shown in Figure \ref{fig:compare} for the emission lines listed in Table \ref{table:lines} (excluding \ion{Fe}{18} 94 \AA , which is discussed in detail in Appendix \ref{sec:Fe18}) that are not used to calculate the DEMs. As with Figure \ref{fig:compare}, we fit these lines in the CHIANTI synthetic spectra resulting from the derived DEMS. The reproduction of these test lines is not expected to show the same level of agreement since they have no impact on the calculated DEMs, but they do reveal interesting trends that lend context to the results.

The \ion{Fe}{8} 131 \AA\ feature is composed of two similar strength \ion{Fe}{8} lines 0.3 \AA\ apart. While it is a weak feature, there is no evidence for any significant contaminating lines within the blended feature. In particular it does not show a response to flares seen in the nearby \ion{Fe}{23} (133 \AA ) feature that would suggest contamination by an unidentified hotter line. It should therefore have a response similar to the \ion{Fe}{8} 168 \AA\ line, but with much lower amplitude. Figure \ref{fig:compare_test} shows that the DEMs reproduce the \ion{Fe}{8} 131 \AA\ feature to about 30\%, with very small variation with density. Given the good reproduction of \ion{Fe}{8} 168 \AA\, this clear trend to poorer agreement with increased flux suggests that there is a non-flare high temperature contribution to the line not included in CHIANTI.

The \ion{Fe}{10} 175 \AA\ line reproduction shows relatively small spreads for the lower densities ($10^{8.5}$ and $10^{9.0}\ \text{cm}^{-3}$), but the trend is depressed about 20\% below the observations. This could indicate a consistent underestimation of the emission measure near log(T)$\sim 6.05$, but the overlapping temperature coverage of \ion{Fe}{9} and \ion{Fe}{11} (Figure \ref{fig:G(T)}) and the excellent reproduction of the \ion{Fe}{11} 188 \AA\ test line (Figure \ref{fig:compare_test}) that was not used in the DEM calculation suggests instead that the CHIANTI database is incomplete in this region of the spectrum.\footnote{We note that the MEGS-A instrument has two slits, A1 and A2, optimized for the 60--180 \AA\ and 160--370 \AA\ wavelength ranges, respectively, and 175 \AA\ is close to the region (away from 171 \AA ) where the A1 and A2 spectra are merged. The responsivity of MEGS-A2 has an edge at 175 \AA , suggesting that this might cause issues for the \ion{Fe}{10} line, but calibration data \citep{Hock2012} show a very smooth transition in the response of the merged spectra at 175 \AA\ and the excellent reproduction of strong lines on either side of this wavelength argue against an instrumental problem.}

As discussed in \S \ref{sec:lines_used} and \ref{sec:density} and shown in Figure \ref{fig:density_dependence}, the \ion{Fe}{13} 202 \AA\ and 204 \AA\ lines have strong and opposite dependencies on density, with the 202 \AA\ line intensity decreasing and 204 \AA\ increasing, respectively, as density increases. The effects of the density sensitivity are obvious in Figure \ref{fig:compare_test} where the reproduced flux in these lines changes as expected with density. As noted in \S \ref{sec:density}, each of these lines suggests a density in the range $10^{8.5}$--$10^{9.0}\ \text{cm}^{-3}$.

The strong \ion{Fe}{15} 284 \AA\ line tends to be over-predicted (between 0\% and 25\% depending on density) during periods of increased activity. This line dominates its region of the spectrum and therefore has very little contamination. We think it unlikely that the calculated DEMs have excess emission measure at the peak response of Fe XV (log(T) $\gtrsim 6.35$) since this temperature is also well sampled by the responses of the strong \ion{Fe}{14} 211 \AA\ and \ion{Fe}{16} 335 \AA\ lines (Figure \ref{fig:G(T)}). It has been suggested that resonance scattering can affect the intensities of strong EUV lines such as \ion{Fe}{15} 284 \AA\ by spatially dispersing photons \citep[e.g.,][]{Schrijver2000,Wood2000}, but \cite{Brickhouse2006} argued that the optical depth of \ion{Fe}{15} 284 \AA\ is unlikely to be high enough and in any case spatial redistribution by resonance scattering should not affect full-Sun irradiance measurements such as those made by EVE. The reconstruction of the \ion{Fe}{15} 284 \AA\ line intensities is consistent with the stated uncertainty.

Overall the test lines demonstrate both the difficulty of this analysis, given its reliance on incomplete EUV emission data, and the robustness of the DEM results to within the stated uncertainty. The results for those lines reproduced most poorly (\ion{Fe}{8}, \ion{Fe}{10}, and \ion{Fe}{13}) can only be explained through systematic effects while the \ion{Fe}{11} and \ion{Fe}{15} lines are reproduce with fidelity similar to the lines used in the DEM calculations.

\section{The Energy and Evolution of the Solar Corona}
\label{sec:evolution}

\begin{figure}[ht] %fig:DEMs
	\centering
	\includegraphics*[trim=0cm 0cm 0cm 0cm, scale=1]{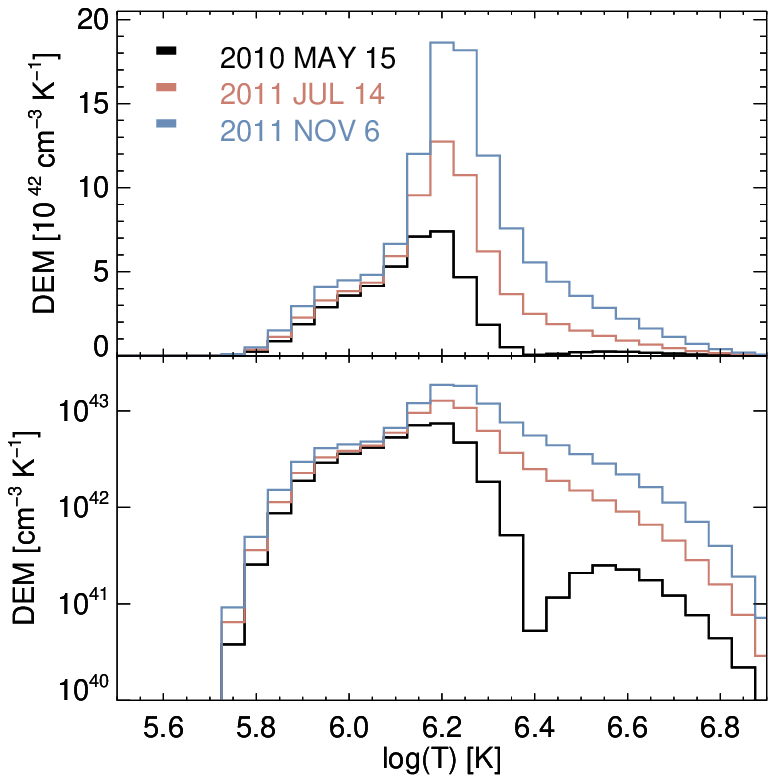}
	\caption{The DEM solutions (with a \textit{top:} linear and \textit{bottom:} logarithmic scale) for a density of $10^{9}\ \text{cm}^{-3}$ at three different activity levels: a solar minimum DEM (black, 2010 May 15); a ``typical'' solar maximum (with frequent B flares) DEM (red, 2011 July 14); and a high activity solar maximum (two M- and and nine C-class flares, all outside the observation window) DEM (blue, 2011 November 6). These show that the DEM remains relatively consistent at temperatures below log(T)=6.1 but that the plasma content of the corona at higher temperatures changes dramatically with solar activity. The dip at log(T)=6.4 in the 2010 May 15 DEM is likely an artifact of the fitting process when the high-temperature lines are weak.}
	\label{fig:DEMs}
\end{figure}

To show quantitatively how the DEMs evolve with solar activity, the DEMs from three different solar activity levels are plotted in Figure \ref{fig:DEMs}. The peak temperature of the DEM is very similar in all cases, just below 1.6 MK (log(T)=6.2) during solar minimum and just above during solar maximum. Additionally, while the low-temperature side of the DEMs are similar on all three dates, there are dramatic differences in the high-temperature side of the DEM. During solar minimum there is very little material above the peak in the temperature distribution, with almost none above 2.5 MK (log(T)=6.4). During solar maximum the bulk of the emission measure lies at temperatures greater than the peak and there is significant emission from plasma up to 6 MK (log(T)=6.8). This compares well with previous work examining the spatial distribution of the DEM \citep{Orlando2001} and the long term evolution of the global DEM \citep{Orlando2004,Argiroffi2008}. These studies used observations from the Yohkoh Soft X-ray Telescope that are sensitive to much higher temperatures than explored here but are less accurate at the low-temperature end of the DEMs that we discuss \citep{Orlando2000}.

\begin{figure*}[ht] %fig:DEM_binned
	\centering
	\includegraphics*[trim=0cm 0cm 0cm 0cm, scale=1]{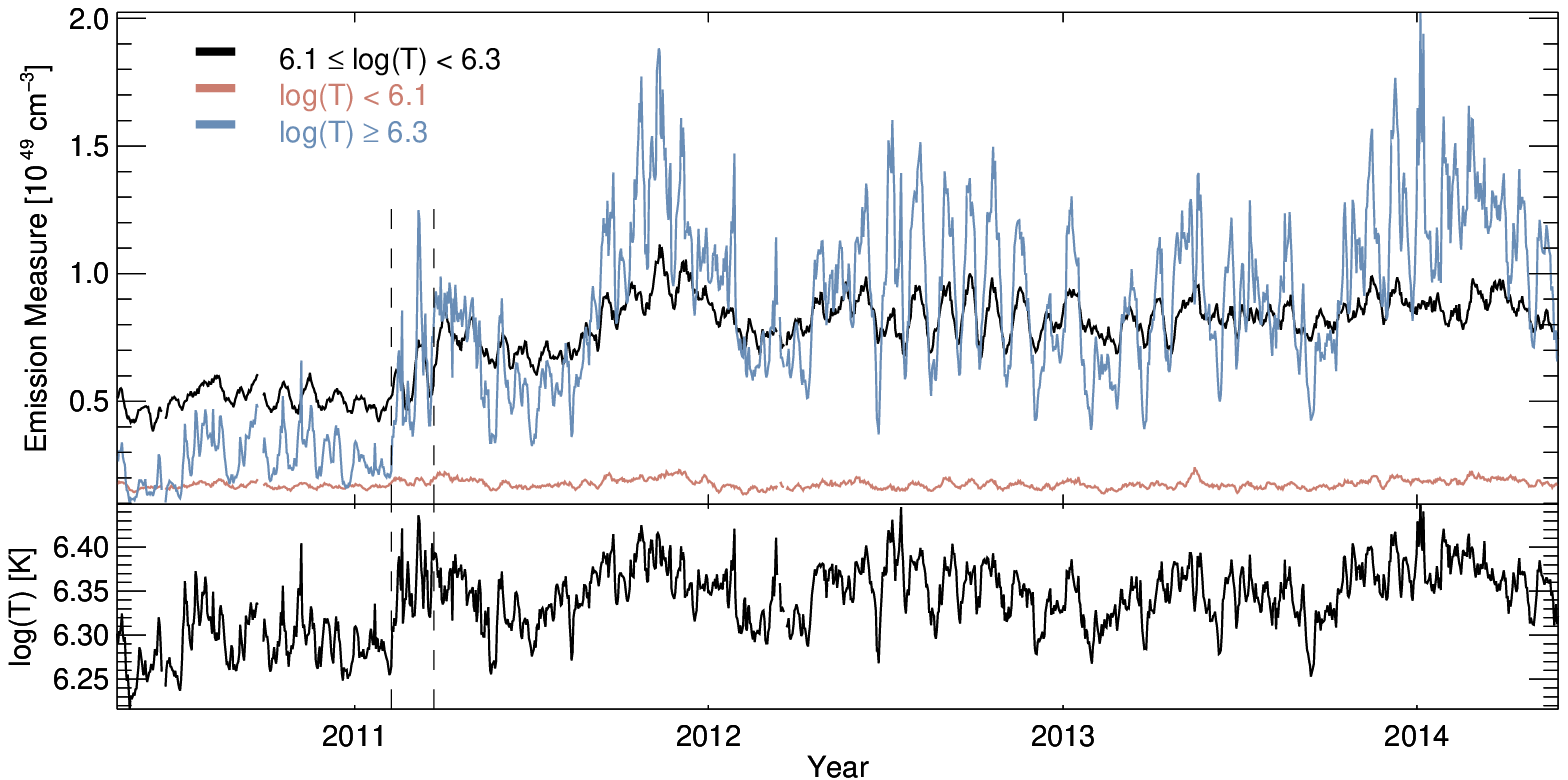}
	\caption{\textit{Top}: the emission measure from Figure \ref{fig:DEM} in the "cool" (red, log(T)$<$ 6.1), "warm" (black, 6.1$\leq$log(T)$<$6.3), and "hot" (blue, log(T)$\geq$ 6.3) temperatures and \textit{bottom}: The emission measure weighted mean temperature for the complete EVE data set. The dashed vertical lines indicate the period of transition from solar minimum to solar maximum discussed in \S \ref{sec:two-state}. This plot shows the consistency of the low temperature corona while the high temperature corona changes dramatically over the solar cycle.}
	\label{fig:DEM_binned}
\end{figure*}

To further illustrate this variation in the DEMs, Figure \ref{fig:DEM_binned} shows the time series of the DEM-weighted average temperature of the solar corona and the DEMs binned into three different temperature ranges: below the temperature peak of the DEM (``cool,'' 5.5$\leq$log(T)$<$6.1), around the temperature peak (``warm,'' 6.1$\leq$log(T)$<$6.3), and above the temperature peak (``hot,'' 6.3$\leq$log(T)$<$6.9). The ``cool'' corona appears to be almost independent of the solar cycle, with little change due either to solar rotation or the activity level over the four years of observation. On the other hand, the ``warm'' and ``hot'' corona vary by a factor of two and an order of magnitude, respectively. These results are consistent with observations that solar activity is manifested primarily through increased hot plasma in active regions, and confirms that there is very little change in the quiet-Sun corona throughout the solar cycle. The increase in high temperature plasma causes the DEM-weighted average temperature to rise from a minimum of 1.6 MK (log(T)=6.2) during solar minimum to above 2.5 MK (log(T)=6.4) during high activity periods at solar maximum.

\begin{figure}[ht] %fig:radiative_loss
	\centering
	\includegraphics*[trim=0cm 0cm 0cm 0cm, scale=1]{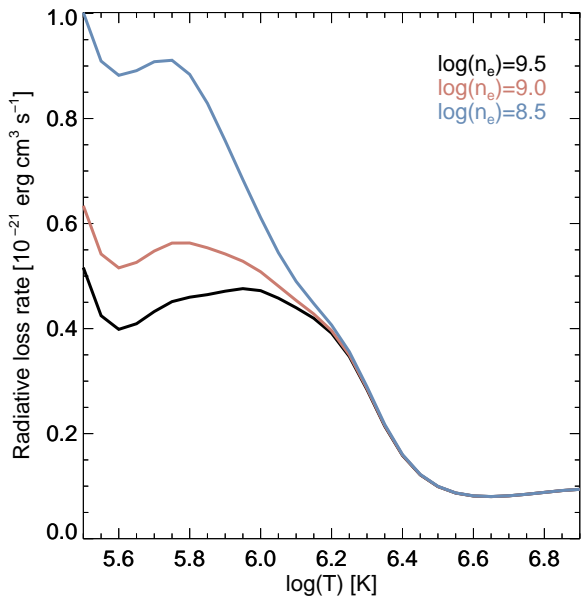}
	\caption{Radiative loss functions per unit volume emission measure as a function of temperature for the three coronal densities considered here. These are calculated using the \texttt{rad\_loss.pro} function in CHIANTI 8.0.2 (see the corresponding plot from CHIANTI 6 in \cite{Dere2009}) assuming \cite{Feldman1992} coronal abundances. This shows that coronal emission is strongly weighted towards the ``cool'' and ``warm'' corona below log(T)=6.3.}
	\label{fig:radiative_loss}
\end{figure}                                                                                                                                                                             

This additional ``hot'' plasma does not appear as additional emission measure cooling through the $<$1~MK range in part due to the temperature dependence of the radiative loss function shown in Figure \ref{fig:radiative_loss}. The loss rate is significantly greater below 1 MK (log(T)=6.0) than at higher temperatures, especially at lower densities. This means that ``hot'' plasma \citep[which experiences significant cooling through conduction to the lower atmosphere, ][]{Antiochos1976,Antiochos1978} will remain so for a long time, and once it drops to sufficiently low temperature it will tend to cool out of the ``warm'' and ``cool'' temperature range quickly. This effect has been termed ``catastrophic cooling" \citep{Reale2012a,Reale2012,Cargill2013} and involves draining of cool plasma back into the lower atmosphere \citep{Bradshaw2010} in addition to radiative cooling.

\subsection{A two-state corona?}
\label{sec:two-state}

The emission from the corona can be described as a combination of the emission from the quiet Sun, coronal holes, and active regions (and trace contributions from smaller features such as filament channels, prominences, etc.). Each of these distinct features have their own characteristic emission spectrum determined by their unique plasma parameters, and the total solar spectrum is the sum of these spectra weighted by their respective covering fractions of the visible solar disk \citep[e.g.,][]{Fontenla2017}. With this description it is clear that the total solar spectrum will change as a function of solar activity as is observed. \textit{A priori} we expect this variation to be continuous as features evolve and rotate on and off the disk and the overall level of activity changes with the solar cycle. However, the data suggest that this is not the case: we observe a rapid transition between the early period of EVE data, near solar minimum conditions, and the later period around solar maximum that suggests a fundamental bifurcation in the DEM over the solar cycle.

\begin{figure*}[ht] %fig:Emission Measure Scatter
	\centering
	\includegraphics*[trim=0cm 0cm 0cm 0cm, width=\linewidth]{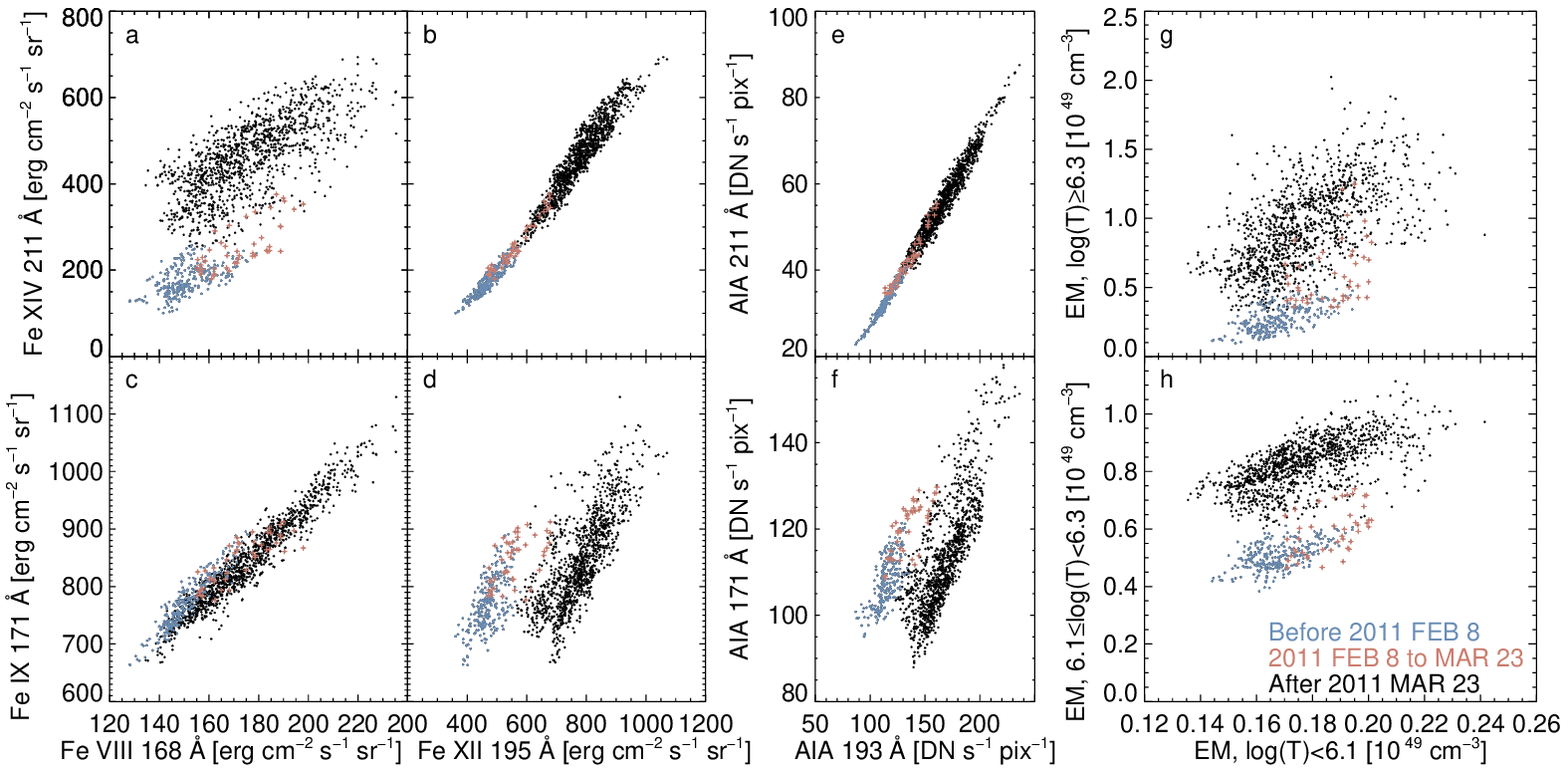}
	\caption{The observed emission in the EVE line pairs: \textit{a}) \ion{Fe}{14} 211 \AA\ vs. \ion{Fe}{8} 168 \AA , \textit{b}) \ion{Fe}{14} 211 \AA\ vs. \ion{Fe}{12} 195 \AA , \textit{c}) \ion{Fe}{9} 171 \AA\ vs. \ion{Fe}{8} 168 \AA , \textit{d}) \ion{Fe}{9} 171 \AA\ vs. \ion{Fe}{12} 195 \AA , the AIA band pairs \textit{e}) 211 \AA\ vs. 193 \AA\ and \textit{f}) 171 \AA\ vs. 193 \AA , and the calculated \textit{g}) ``hot'' vs. ``cool'' and \textit{h}) ``warm'' vs. ``cool'' emission measure. In all plots, the blue points indicate solar minimum (before 2011 February 8) and the black points indicate solar maximum data (after 2011 March 23) while the red crosses indicate measurements during the intervening transition (see also Figures \ref{fig:EM} and \ref{fig:DEM_binned}). Panes \textit{b} and \textit{c} compare lines originating from similar temperature plasma that have a simple linear relationship. Panes \textit{a} and \textit{d} compare lines originating from different temperature plasma with a clear difference between the solar minimum and maximum trends. Panes \textit{e} and \textit{f} are equivalent to panes \textit{b} and \textit{d}, respectively, but with the disk-integrated AIA observations. The AIA measurements are the mean values derived from the AIA ``light curve" data files provided in the SDO ``SunInTime" daily directories. Panes \textit{g} and \textit{h} demonstrate that the ``cool'' plasma is nearly decoupled from the ``warm'' and ``hot'' plasma with different solar minimum and maximum trends.}
	\label{fig:EM_scatter}
\end{figure*}                                                                                                                                                                             

This discontinuity is shown in Figure \ref{fig:EM_scatter} for four sets of observed line fluxes, three AIA bands, and the calculated DEMs. Panels \textit{a}--\textit{d} show the relationship over the four years of EVE data between the \ion{Fe}{14} 211 \AA\ and \ion{Fe}{9} 171 \AA , and \ion{Fe}{8} 168 \AA\ and \ion{Fe}{12} 195 \AA\ lines. Both \ion{Fe}{8} 168 \AA\ and \ion{Fe}{9} 171 \AA\ have their strongest responses at temperatures below the DEM peak, whereas \ion{Fe}{12} 195 \AA\ and \ion{Fe}{14} 211 \AA\ contribute at or above the temperature peak (see Figure \ref{fig:G(T)}). Panels \textit{a} and \textit{d} show general linear trends between the line pairs that cluster into two distinct sets, one for the solar minimum conditions before 2011 February 8 (blue points) and one for the solar maximum conditions after 2011 March 23 (black points), although these dates are chosen somewhat arbitrarily. For example, a given observed flux in \ion{Fe}{9} 171 \AA\ implies two very different \ion{Fe}{12} 195 \AA\ fluxes, depending on the level of solar activity. The most obvious explanation for such a sharp transition in the observed fluxes would be a calibration error in the EVE MEGS-A data. However, while the calibration of MEGS-A spectra is updated with rocket under-flights \citep[including one on 2011 March 23, near the division between solar minimum and maximum identified here,][]{Hock2012}, these calibrations are applied in a continuous fashion specifically designed to prevent the kind of discontinuity observed here. Panels \textit{b} and \textit{c} show that for lines originating from plasma of similar temperature, the linear trends are uniform, with the transition points (red crosses) clearly connecting the solar minimum and maximum trends. The fact that this activity discontinuity is seen across multiple, but not all, line pairs strongly suggests that it is a true feature of the emission and not a result of calibration errors. Additionally, panes \textit{e} and \textit{f} compare observations from the Atmospheric Imaging Assembly (AIA, also on SDO) and are nearly identical to their EVE counterparts in panes \textit{b} and \textit{d}.

The same discontinuity appears in panels \textit{g} and \textit{h} of Figure \ref{fig:EM_scatter} which show a similar linear relationship and coronal activity clustering but for the total ``hot" v.s ``cool" and ``warm" vs ``cool" emission measure, respectively. This indicates that the shape of the DEM changes discontinuously between solar minimum and solar maximum, with almost no increase in the ``cool" plasma after activity turns on. If the three sets of points formed a single linear feature with a gap during the transition (like panel \textit{b}), it would simply indicate a rapid turn-on of activity; instead, the fact that the black and blue sets of points both have similar slopes but are offset relative to one another appears to indicate a fundamental change in the shape of the DEM. We note that the timing of this transition period in 2011 February--March is of interest because the first X-class flare of cycle 24 occurred on 2011 February 15, during the transition period. The correlation of the change in coronal behavior with other solar properties, and magnetic field characteristics in particular, will be addressed in a future paper.

\subsection{The coronal thermal energy content}
\label{sec:energy_evolution}

The total thermal content of the corona is of interest for understanding the energetics of the solar atmosphere and the role of heat transfer in the temperature structure of the corona. To our knowledge, this quantity has not previously been addressed in any detail. We can use EVE DEMs derived here to estimate the coronal thermal content.

The dominant components of the corona are protons, alpha particles, and electrons. Accordingly, we can express the thermal energy as:

\begin{align}
	\text{E} &= \frac{3}{2}\int\displaylimits_{\text{V}}\int\frac{d}{d\text{T}}\bigl(\text{n}_{\text{e}}(\text{T})+\text{n}_{\text{H}}(\text{T})+\text{n}_{\text{He}}(\text{T})\bigr)\times \nonumber\\
	& \hspace{2cm} \text{k}_{\text{B}}\text{T}\ d\text{T}\ d\text{V} \nonumber\\
	&= 3.375\ \text{k}_{\text{B}}\int\displaylimits_{\text{V}}\int\ \left(\frac{d}{d\text{T}}\ \text{n}_{\text{H}}(\text{T})\right) \text{T}\ d\text{T}\ d\text{V}
	\label{eqn:energy}
\end{align}
where k$_{\text{B}}$ is Boltzmann's constant, we assume n$_{\text{e}}$ = n$_{\text{H}}$ + 2n$_{\text{He}}$ in a fully-ionized corona, and adopt the standard value n$_{\text{He}}$/n$_{\text{H}}$=0.085 \citep{Asplund2009}. Using Equation \ref{eqn:DEM} and noting that we assume a constant coronal density, we can relate the total energy to the DEM by:

\begin{equation}
	\text{E} = \frac{3.375\ \text{k}_{\text{B}}}{\text{n}_{\text{e}}}\int\ \text{DEM(T)}\ \text{T}\ d\text{T}
\label{eqn:energy_DEM}
\end{equation}
A large uncertainty arises from the division by n$_{\text{e}}$ in Equation \ref{eqn:energy_DEM}. Since the DEM is density-squared and Figure \ref{fig:EM} shows that the density assumption has only a small effect on the derived DEM, the calculated coronal energy is essentially inversely proportional to the density used in the DEM calculation. Thus, for the assumed density range $10^{8.5}$--$10^{9.5}\ \text{cm}^{-3}$, the energy can vary by a factor of about three from the value obtained using the central $10^{9}\ \text{cm}^{-3}$.

Assuming a constant density for this energy calculation is fundamentally different from the constant density assumption made in \S \ref{sec:density} where an order of magnitude change in density typically caused only a 50\% change in emission. Here, the constant density assumption allows us to pull the density out of the integral and is equivalent to assuming that the DEM results only from variations in the emitting volume with temperature. This means that all the caveats mentioned in \S \ref{sec:density} relating to a spatially and temporally variable coronal density can have an even larger distorting effect when calculating the coronal thermal energy. Nonetheless, this approach yields an order-of-magnitude estimate of coronal energy that is useful for discussing trends with solar activity.

\begin{figure*}[ht] %fig:energy
	\centering
	\includegraphics*[trim=0cm 0cm 0cm 0cm, scale=1]{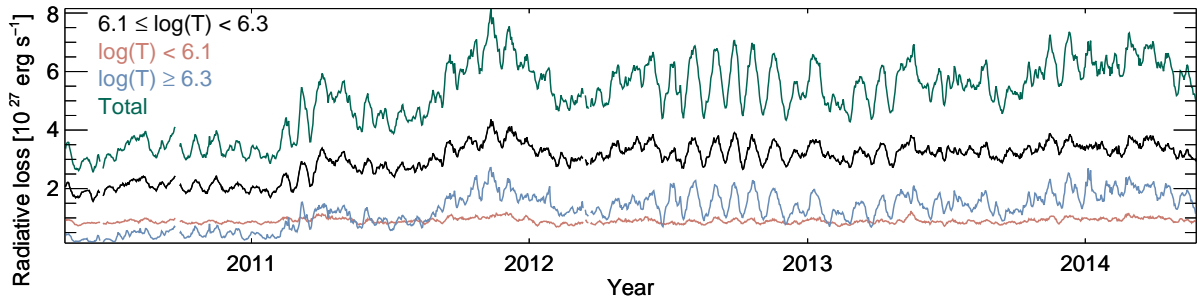}
	\caption{The total radiative energy loss rate from the corona derived from the EVE DEMs assuming a density of $10^{9}\ \text{cm}^{-3}$ and coronal abundances. The total energy loss rate is shown in green and the contributions from plasma in the temperature ranges from Figure \ref{fig:DEM_binned} are also plotted: ``cool'' (red, log(T)$<$6.1), ``warm'' (black, 6.1$\leq$log(T)$<$6.3), and ``hot'' (blue, log(T)$\geq$6.3).}
	\label{fig:energy}
\end{figure*}                                                                                                                                                                             

We find that the visible coronal volume contains on the order of $10^{31}\ \text{erg}$ of thermal energy, suggesting that the total coronal volume contains less energy than a typical X-class solar flare \citep{Sun2012a,Tziotziou2013,Aschwanden2014}. This thermal energy increases by just under an order of magnitude at periods of peak activity, compared to the low-activity levels early in the SDO mission, both due to the general increase in emission measure as well as the specific increase in ``hot" plasma seen in Figure \ref{fig:DEM_binned}. Due to the factor of T inside the integral in Equation \ref{eqn:energy_DEM} the ``hot" corona disproportionately influences the total thermal energy, containing the majority of the energy during solar maximum. Conversely, the ``cool" corona contains only a very small fraction of the thermal energy, even at low activity levels.

Total radiative energy output from the corona is calculated by integrating the product of the DEM with the radiative loss curve for $10^{9}\ \text{cm}^{-3}$ shown in Figure \ref{fig:radiative_loss}. This result is shown in Figure \ref{fig:energy} with the contributions from the three temperature regimes discussed in \S \ref{sec:evolution} plotted individually. It is striking that the total radiative energy loss varies by only a factor of three over the wide range of coronal conditions observed in the four-year period. This results from the shape of the radiative loss curve and the fact that radiation is much more efficient from plasma with temperatures below 2 MK (log(T)=6.3) than from hotter plasma. This means that most of the emission results from the relatively low variability ``cool'' and ``warm'' components even though they contain the minority of the energy. If we assume that the radiative loss is from a single hemisphere (even though a fraction of visible off-limb plasma will always be beyond the solar limb and therefore above the far hemisphere) we can calculate the hemisphere-averaged coronal radiative energy loss. The typical $3 \times 10^{27}\ \text{erg s}^{-1}$ solar minimum rate from Figure \ref{fig:energy} corresponds to an average radiative flux of $1 \times 10^{5}\ \text{erg cm}^{-2}\ \text{s}^{-1}$, exactly matching the traditional estimate for quiet-Sun regions. A typical solar maximum value of $6 \times 10^{27}\ \text{erg s}^{-1}$ corresponds to $2 \times 10^{5}\ \text{erg cm}^{-2}\ \text{s}^{-1}$, well below the typical radiative loss rate of an individual active region, $5 \times 10^{6}\ \text{erg cm}^{-2}\ \text{s}^{-1}$ \citep[e.g.,][]{Withbroe1977}.

Dividing the total energy by the radiative energy loss rate produces the coronal energy turnover timescale, the time needed to radiate away the total energy at the calculated loss rate. The resulting timescale is about an hour, and typically longer during solar maximum than solar minimum. The timescale is shorter during solar minimum conditions both because the total coronal energy is lower and because the ``cool'' and ``warm'' components, which radiate more rapidly, contain a larger fraction of the total energy. This timescale only accounts for radiative losses and does not include heat conduction into the lower (and cooler) solar atmosphere, meaning that the actual energy replacement timescale for the solar corona will be significantly shorter than estimated here \citep[e.g.][]{Rosner1978, Klimchuk2008}. This is a characteristic timescale for the global corona; as discussed in \S \ref{sec:density} the density variation between environments in the corona (coronal holes, quiet Sun, active regions, etc.) will result in greatly varying energy replenishment times in different coronal features.

\section{Conclusion}
\label{sec:conclusion}

We have used EVE median spectra to generate daily DEM distributions for the entire four-year period of operation of the EVE MEGS-A detector. This resulted in DEMs derived from a uniform data set beginning in 2010 at near-solar-minimum conditions and continuing through the maximum of solar cycle 24. The DEMs are calculated using six strong line features dominated by Fe lines of charge states \ion{}{8}, \ion{}{9}, \ion{}{11}, \ion{}{12}, \ion{}{14}, and \ion{}{16} that adequately sample the quiet-Sun coronal temperature range 0.3 to 5 MK (log(T)=5.5-6.7). We investigated other strong lines and found them to lead to poorer DEM solutions. In particular, we demonstrated (see Appendix \ref{sec:Fe18}) that CHIANTI does not currently reproduce EVE spectra in the wavelength range near the \ion{Fe}{18} line at 93.9 \AA , making it unsuitable as a constraint on high-temperature quiet-Sun emission.

In order to generate the temperature responses, a quiet-Sun coronal abundance for Fe and density have to be specified. We used the standard \cite{Feldman1992} coronal Fe abundance and a density of $10^{9.0}\ \text{cm}^{-3}$, with the results for $10^{8.5}$ and $10^{9.5}\ \text{cm}^{-3}$ serving as a measure in the uncertainty in this choice. The short term daily variability, uncertainties in the atomic data, calibration of EVE, and spectral fitting also contributed to the overall uncertainty in our results, estimated to be no better than 15\%. Future improvements in the relevant atomic data and better understanding of coronal abundances will alter our results. We therefore regard the trends evident in our DEM results to be more robust than their absolute values.

The behavior of the coronal DEM over the four-year period is consistent with an intuitive understanding of a corona consisting of two primary components: the quiet Sun, and active regions. The quiet Sun DEM component with a peak temperature of 1.6 MK (log(T)=6.2) and little emission measure above 2 MK (log(T)=6.3) is present and relatively constant throughout the solar cycle. This suggests that, outside of active regions, there is little difference in the quiet Sun between solar minimum and solar maximum. The active region DEM component with a peak temperature above 2 MK (log(T)=6.3) varies by more than an order of magnitude with the solar cycle. Plasma in the 1.25--2 MK (log(T)=6.1--6.3) range varies by a factor of three over the four years and alternates with the hotter component as to which is (quantitatively) dominant during solar maximum.

We estimated the total energy of the visible solar corona, its radiative energy loss rate, and the corresponding energy turnover timescale. During solar maximum, the higher-temperature component dominates the energy content of the corona. The coronal radiative energy loss rate varies by only a factor of three over the solar cycle, due to the fact that the more stable cooler coronal material has a loss rate much higher than the highly variable ``hot'' component. The energy turnover timescale is on the order of an hour, but results for the total energy and the energy turnover timescale are very uncertain due to the strong dependence of both the total energy and the turnover timescale on density. Additionally, we identified a discontinuity in the behavior of coronal diagnostics in 2011 February--March, around the time of the first X-class flare of cycle 24, that suggests fundamentally different behavior in the corona under solar minimum and maximum conditions.

The DEMs derived here will be used in a subsequent paper (Schonfeld et al. 2017, in prep) to discuss the evolution of the relationship between the solar F$_{10.7}$ index \citep[e.g.,][]{Tapping2013} and the coronal ionizing radiation for which it serves as a proxy in terrestrial atmospheric models, as well as the correspondence with the evolution of global solar magnetic fields over the solar cycle.

\acknowledgements
\textit{Acknowledgements:} Data supplied courtesy of the SDO/EVE consortium. SDO is the first mission to be launched for NASA's Living With a Star (LWS) Program. CHIANTI is a collaborative project involving George Mason University, the University of Michigan (USA) and the University of Cambridge (UK). This research has been made possible with funding from AFOSR LRIR 14RV14COR and 17RVCOR416, and FA9550-15-1-0014, NSF Career Award \#1255024, and PAARE NSF:0849986. We thank Carl Henney and the anonymous reviewer for valuable comments on the manuscript.

\appendix

\section{The EVE Spectrum Around Fe XVIII 93.9 \AA}
\label{sec:Fe18}

EVE data for the \ion{Fe}{18} 93.9 \AA\ line is a commonly used diagnostic of solar flares because it is one of the strongest hot lines observed by EVE. With a peak emission temperature of 7 MK (log(T)=6.85), it is an ideal flare diagnostic, with small response to typical coronal temperatures but easily reached by even small flares \citep[e.g.,][]{Warren2011,Petkaki2012}. This region of the spectrum is especially notable because the Atmospheric Imaging Assembly on SDO \citep[AIA;][]{Lemen2012} employs 94 \AA -bandpass images as one of the primary diagnostics of high temperature coronal plasma. However, this wavelength range lacks well-calibrated high resolution spectra of the quality that is available at longer EUV wavelengths and this limits the line identifications available for the CHIANTI database. The NASA Extreme Ultraviolet Explorer (EUVE) mission observed the 93.9 \AA\ Fe XVIII line in a large number of active stars, but with insufficient signal-to-noise to identify cooler lines at neighbouring wavelengths \citep[e.g.,][]{Mewe1995,Sanz-Forcada2003}. Such cool (e.g., Fe \textsc{X}) lines were known to lie near the Fe \textsc{XVIII} line when SDO was launched \citep{Boerner2012}, but CHIANTI did not reproduce the spectrum completely \citep{Aschwanden2011, Reale2011, Testa2012, Aschwanden2013a}, although it is believed that the \ion{Fe}{18} 93.9 \AA\ line itself is correctly represented in CHIANTI \citep{Warren2012,DelZanna2013b}. \cite{DelZanna2012} identified an \ion{Fe}{14} line that is blended with \ion{Fe}{18} in the EVE spectra, and empirical corrections have been made to the AIA temperature response functions \citep{DelZanna2013b, Boerner2014}. These complications are often avoidable in flare studies where the pre-flare emission can be subtracted \citep[e.g.][]{Warren2013a} to isolate only contributions from the high-temperature flare plasma which emits primarily in the well-characterized \ion{Fe}{18} line.

\begin{figure}[ht] %fig:Fe18_spectra
	\centering
	\includegraphics*[trim=0cm 0cm 0cm 0cm, scale=1]{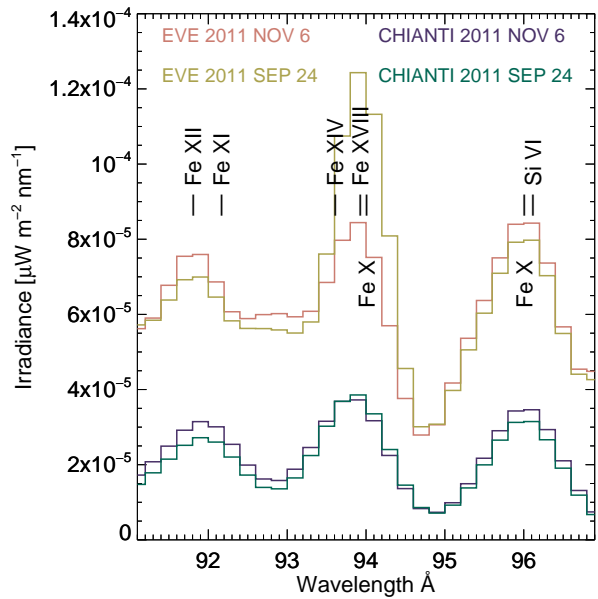}
	\caption{The EVE MEGS-A spectra and CHIANTI model spectra calculated from the DEMs derived as described in \S \ref{sec:DEM} of the wavelength range surrounding the \ion{Fe}{18} 93.9 \AA\ line with the strongest emission lines marked. The 2011 September 24 spectrum includes emission from a flare while the 2012 November 6 spectrum is typical of non-flaring periods during solar maximum. The CHIANTI models significantly underestimate the emission across this wavelength range. The synthetic spectrum fails to reproduce the increased flare emission on 2011 September 24 because none of the lines used in the DEM calculation are sensitive to flare emission (Figure \ref{fig:G(T)}).}
	\label{fig:Fe18_spectra}
\end{figure}                                                                                                                                                                             

Because our analysis focuses specifically on the non-flaring corona, this type of subtraction technique is inappropriate. We therefore investigate the spectrum surrounding the \ion{Fe}{18} 93.9 \AA\ line to determine whether it contributes sufficiently to the EVE spectra to constrain high temperature emission. Figure \ref{fig:Fe18_spectra} compares EVE spectra with Version 8.0.2 of the CHIANTI model spectra generated using DEMs computed as described in \S \ref{sec:DEM}. The CHIANTI models show the same three peaks visible in the EVE data, but with amplitudes about half or less of what is observed in the 91--97 \AA\ range. Because these emission features span a wide range of coronal temperatures (1--7 MK, log(T)=6--6.85), including those well represented in the calculated DEMs, it is clear that some other factor is affecting this region of the spectrum. The three most likely explanations are: problems with the EVE MEGS-A calibration in this region of the spectrum, unexplained continuum emission, or significant emission from lines not identified in CHIANTI.

\begin{figure*}[ht] %fig:Fe18_index
	\centering
	\includegraphics*[trim=0cm 0cm 0cm 0cm, scale=1]{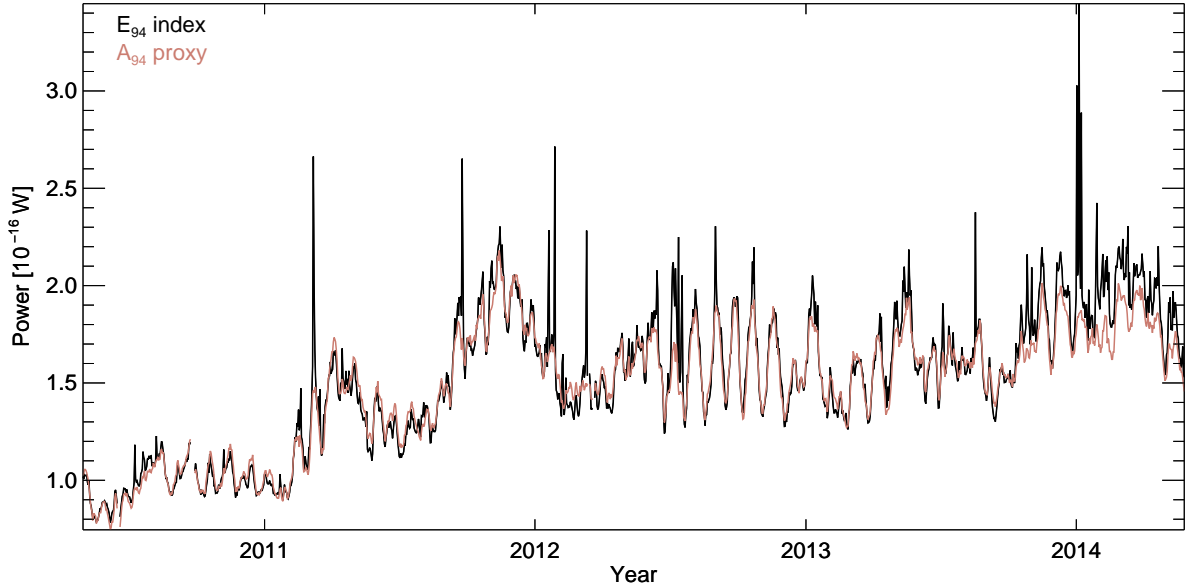}
	\caption{The time series of the intensity in the AIA 94 \AA\ EVE index $E_{94}$ (black) and the 171/211 \AA\ proxy $A_{94}$ (red) as observed by MEGS-A. The sharp spikes in $E_{94}$ where it deviates from $A_{94}$ are times when significant \ion{Fe}{18} emission contributes to the 93.9 \AA\ line suggesting the presence of significant flares with high temperature plasma.}
	\label{fig:Fe18_index}
\end{figure*}                                                                                                                                                                            

This conclusion does not address the issue of whether EVE daily non-flare spectra contain significant \ion{Fe}{18} emission. A number of methods for isolating \ion{Fe}{18} emission from AIA 94 \AA\ images have been developed \citep[e.g.,][]{Warren2012,DelZanna2013b}. We find an analogous method for EVE spectra. We define EVE indices for the AIA 94, 171 and 211 \AA\ bandpasses by integrating over the product of the daily EVE median spectra $I^{EVE} (\lambda)$ with the AIA effective area functions of wavelength\footnote{Derived from the Version 6 AIA response files in the SolarSoft distribution of AIA software \citep{Boerner2012}. The unit of effective area is cm$^2$.}, $R^{AIA} (\lambda)$, for these windows and summing over wavelength:

\begin{align}
	E_{94} \,&=\, \int \, I^{EVE}(\lambda)\ R^{AIA}_{94} (\lambda)\ d\lambda\nonumber \\
	E_{171} \,&=\, \int \, I^{EVE}(\lambda)\ R^{AIA}_{171} (\lambda)\ d\lambda\nonumber \\
	E_{211} \,&=\, \int \, I^{EVE}(\lambda)\ R^{AIA}_{211} (\lambda)\ d\lambda\nonumber \\
	\label{eqn:EVE_indices}
\end{align}
Since the units of EVE irradiance are W m$^{-2}$ nm$^{-1}$ and we sum over wavelength and multiply by effective area, these indices have units of W (nominally, the power received by each AIA detector). The following expression proves to be a surprisingly good proxy for $E_{94}$:

\begin{equation}
	A_{94} \,=\, 0.0235\  {{E_{171}\ E_{211}} \over {E_{171}\,+\,E_{211}}}
\end{equation}
Figure \ref{fig:Fe18_index} compares $E_{94}$ (black line) with $A_{94}$ (red line) for the period of MEGS-A observations. $A_{94}$ is generally within a few percent of the EVE index on all days except for a limited number of days when there are sharp spikes in $E_{94}$. The ability of AIA 171 and 193 \AA\ bandpasses to reproduce the 94 \AA\ behavior is not surprising. This is because the 94 \AA\ region contains \ion{Fe}{10} (94.0 \AA ) and \ion{Fe}{14} (93.2 and 93.6 \AA ) lines in addition to \ion{Fe}{18} while the 171 \AA\ AIA bandpass includes \ion{Fe}{9} with a temperature similar to \ion{Fe}{10} and the 211 \AA\ AIA bandpass is dominated by \ion{Fe}{14}. Neither of these bandpasses contains any significant lines hotter than \ion{Fe}{14}. The most widespread proxy used to separate \ion{Fe}{18} from the AIA images also uses the 171 and 211 \AA\ images but in a linear combination (with two free parameters) proposed by \cite{DelZanna2013b}, while \cite{Warren2012} used a polynomial combination of 171 and 193 \AA\ with seven free parameters and \cite{Reale2011} suggested just the AIA 171 \AA\ data to estimate the cool contribution to 94 \AA .

Investigation of solar activity on days when the $A_{94}$ proxy departs significantly from the EVE $E_{94}$ index shows that they are all days when significant, usually long-duration, flaring occurs in the 19--21 UT window used to derive the EVE median spectra. On this basis, we argue that it is likely that the EVE full-Sun spectrum around 94 \AA\ only contains significant \ion{Fe}{18} emission when flares contribute, and that EVE data do not provide evidence for significant \ion{Fe}{18} emission in non-flaring full-Sun spectra. The DEMs we derive from the EVE data do not suggest the presence of \ion{Fe}{18} emission down to the 7\% MEGS-A precision, so we conclude that EVE spectra at 94 \AA\ do not help constrain the high-temperature emission from the quiet Sun. Imaging observations that better isolate hot areas in active regions will be more successful in constraining the hot component of the solar corona since they are not competing with the cool emission from the entire Sun, as is the case for EVE data.

\begin{table*}[ht] %table:lines
	\begin{center}
		\caption{Analyzed EUV mission lines}
		\begin{tabular}{|l|c|c|c|c|c|}
			\hline
			\T Ion & \hspace{1cm}Wavelength [\AA]\hspace{1cm} & Peak [log(T)] & Relative G(T) & Lower State & Upper State \B \\ 
			\hline
			
			\T Fe VIII & 131.2400 & 5.75 & $2.372\times 10^{-25}, 0.047$ & 
				$3\text{s}^{2}\ 3\text{p}^{6}\ 3\text{d}\  {^{2}}\text{D}_{5/2}$ & 
				$3\text{s}^{2}\ 3\text{p}^{6}\ 4\text{f}\ {^{2}}\text{F}_{7/2}$ \\
			\hspace{0.5cm} Fe VIII & 130.9410 & 5.75 & $0.668$ & 
				$3\text{s}^{2}\ 3\text{p}^{6}\ 3\text{d}\ {^{2}}\text{D}_{3/2}$ & 
				$3\text{s}^{2}\ 3\text{p}^{6}\ 4\text{f}\ {^{2}}\text{F}_{5/2}$ \B \\
			\hline			
			
			\T \textbf{Fe VIII} & \textbf{168.1720} & 5.75 & $1.315\times10^{-24}, 0.260$ & 
				$3\text{s}^{2}\ 3\text{p}^{6}\ 3\text{d}\  {^{2}}\text{D}_{5/2}$ & 
				$3\text{s}^{2}\ 3\text{p}^{5}\ 3\text{d}^{2}\ {^{2}}\text{D}_{5/2}$ \\
			\hspace{0.5cm} Fe VIII & 167.4860 & 5.75 & $0.626$ & 
				$3\text{s}^{2}\ 3\text{p}^{6}\ 3\text{d}\ {^{2}}\text{D}_{3/2}$ & 
				$3\text{s}^{2}\ 3\text{p}^{5}\ 3\text{d}^2\ {^{2}}\text{D}_{3/2}$ \\
			\hspace{0.5cm} Fe VIII & 167.6540 & 5.75 & $0.060$ & 
				$3\text{s}^{2}\ 3\text{p}^{6}\ 3\text{d}\ {^{2}}\text{D}_{3/2}$ & 
				$3\text{s}^{2}\ 3\text{p}^{5}\ 3\text{d}^2\ {^{2}}\text{D}_{5/2}$ \\
			\hspace{0.5cm} Fe VIII & 168.0030 & 5.75 & $0.051$ & 
				$3\text{s}^{2}\ 3\text{p}^{6}\ 3\text{d}\ {^{2}}\text{D}_{5/2}$ & 
				$3\text{s}^{2}\ 3\text{p}^{5}\ 3\text{d}^2\ {^{2}}\text{D}_{3/2}$ \\
			\hspace{0.5cm} Fe VIII & 168.5440 & 5.75 & $0.599$ &
				$3\text{s}^{2}\ 3\text{p}^{6}\ 3\text{d}\  {^{2}}\text{D}_{5/2}$ &
				$3\text{s}^{2}\ 3\text{p}^{5}\ 3\text{d}^{2}\ {^{2}}\text{P}_{3/2}$ \\
			\hspace{0.5cm} Fe VIII & 168.9290 & 5.75 & $0.312$ & 
				$3\text{s}^{2}\ 3\text{p}^{6}\ 3\text{d}\ {^{2}}\text{D}_{3/2}$ & 
				$3\text{s}^{2}\ 3\text{p}^{5}\ 3\text{d}^2\ {^{2}}\text{P}_{1/2}$ \B \\
			\hline
			
			\T \textbf{Fe IX} & \textbf{171.0730} & 5.95 & $5.048\times 10^{-24}, 1.000$ & 
				$3\text{s}^{2}\ 3\text{p}^{6}\ {^{1}}\text{S}_{0}$ & 
				$3\text{s}^{2}\ 3\text{p}^{5}\ 3\text{d}\ {^{1}}\text{P}_{1}$ \B \\
			\hline
			
			\T Fe X & 174.5310 & 6.05 & $2.348\times 10^{-24}, 0.465$ & 
				$3\text{s}^{2}\ 3\text{p}^{5}\ {^{2}}\text{P}_{3/2}$ & 
				$3\text{s}^{2}\ 3\text{p}^{4}\ 3\text{d}\ {^{2}}\text{D}_{5/2}$ \B \\
			\hline
			
			\T \textbf{Fe XI} & \textbf{180.4010} & 6.15 & $1.760\times 10^{-24}, 0.349$ &
				$3\text{s}^{2}\ 3\text{p}^{4}\ {^{3}}\text{P}_{2}$ &
				$3\text{s}^{2}\ 3\text{p}^{3}\ 3\text{d}\ {^{3}}\text{D}_{3}$ \\
			\hspace{0.5cm} Fe X & 180.4410 & 6.05 & $0.106$ &
				$3\text{s}^{2}\ 3\text{p}^{5}\ {^{2}}\text{P}_{1/2}$ &
				$3\text{s}^{2}\ 3\text{p}^{4}\ 3\text{d}\ {^{2}}\text{P}_{1/2}$ \B \\
			\hline
			
			\T Fe XI & 188.2160 & 6.15 & $8.619\times 10^{-25}, 0.171$ &
				$3\text{s}^{2}\ 3\text{p}^{4}\ {^{3}}\text{P}_{2}$ &
				$3\text{s}^{2}\ 3\text{p}^{3}\ 3\text{d}\ {^{3}}\text{P}_{2}$ \\
			\hspace{0.5cm} Fe XI & 188.2990 & 6.15 & $0.602$ &
				$3\text{s}^{2}\ 3\text{p}^{4}\ {^{3}}\text{P}_{2}$ &
				$3\text{s}^{2}\ 3\text{p}^{3}\ 3\text{d}\ {^{1}}\text{P}_{1}$ \\
			\hspace{0.5cm} Fe IX & 188.4930 & 5.95 & $0.277$ &
				$3\text{s}^{2}\ 3\text{p}^{5}\ 3\text{d}\ {^{3}}\text{F}_{4}$ &
				$3\text{s}^{2}\ 3\text{p}^{4}\ 3\text{d}^{2}\ {^{3}}\text{G}_{5}$ \B \\
			\hline			
			
			\T \textbf{Fe XII} & \textbf{195.1190} & 6.20 & $1.298\times10^{-24}, 0.257$ &
				$3\text{s}^{2}\ 3\text{p}^{3}\ {^{4}}\text{S}_{3/2}$ &
				$3\text{s}^{2}\ 3\text{p}^{2}\ 3\text{d}\ {^{4}}\text{P}_{5/2}$ \B \\
			\hline
			
			\T Fe XIII & 202.0440 & 6.25 & $6.936\times10^{-25}, 0.137$ &
				$3\text{s}^{2}\ 3\text{p}^{2}\ {^{3}}\text{P}_{0}$ &
				$3\text{s}^{2}\ 3\text{p}\ 3\text{d}\ {^{3}}\text{P}_{1}$ \\
			\hspace{0.5cm} Fe XI & 201.7340 & 6.15 & $0.091$ &
				$3\text{s}^{2}\ 3\text{p}^{4}\ {^{1}}\text{D}_{2}$ &
				$3\text{s}^{2}\ 3\text{p}^{3}\ 3\text{d}\ {^{3}}\text{S}_{1}$ \\
			\hspace{0.5cm} Fe XI & 202.4240 & 6.15 & $0.097$ &
				$3\text{s}^{2}\ 3\text{p}^{4}\ {^{3}}\text{P}_{2}$ &
				$3\text{s}^{2}\ 3\text{p}^{3}\ 3\text{d}\ {^{3}}\text{P}_{2}$ \B \\
			\hline
			
			\T Fe XIII & 203.8260 & 6.25 & $5.251\times10^{-25}, 0.104$ &
				$3\text{s}^{2}\ 3\text{p}^{2}\ {^{3}}\text{P}_{2}$ &
				$3\text{s}^{2}\ 3\text{p}\ 3\text{d}\ {^{3}}\text{D}_{3}$ \\
			\hspace{0.5cm} Fe XII & 203.7280 & 6.20 & $0.201$ &
				$3\text{s}^{2}\ 3\text{p}^{3}\ {^{2}}\text{D}_{5/2}$ &
				$3\text{s}^{2}\ 3\text{p}^{2}\ 3\text{d}\ {^{2}}\text{D}_{5/2}$ \\
			\hspace{0.5cm} Fe XIII & 203.7950 & 6.25 & $0.402$ &
				$3\text{s}^{2}\ 3\text{p}^{2}\ {^{3}}\text{P}_{2}$ &
				$3\text{s}^{2}\ 3\text{p}\ 3\text{d}\ {^{3}}\text{D}_{2}$ \\
			\hspace{0.5cm} Fe XIII & 204.2620 & 6.25 & $0.125$ &
				$3\text{s}^{2}\ 3\text{p}^{2}\ {^{3}}\text{P}_{1}$ &
				$3\text{s}^{2}\ 3\text{p}\ 3\text{d}\ {^{1}}\text{D}_{2}$ \B \\
			\hline
			
			\T \textbf{Fe XIV} & \textbf{211.3172} & 6.30 & $9.191\times 10^{-25}, 0.182$ &
				$3\text{s}^{2}\ 3\text{p}\ {^{2}}\text{P}_{1/2}$ &
				$3\text{s}^{2}\ 3\text{d}\ {^{2}}\text{D}_{3/2}$ \B \\
			\hline
			
			\T Fe XV & 284.1630 & 6.35 & $2.518\times 10^{-24}, 0.499$ &
				$3\text{s}^{2}\ {^{1}}\text{S}_{0}$ & 
				$3\text{s}\ 3\text{p}\ {^{1}}\text{P}_{1}$ \B \\
			\hline			
			
			\T \textbf{Fe XVI} & \textbf{335.4090} & 6.45 & $1.123\times 10^{-24}, 0.222$ &
				$3\text{s}\ {^{2}}\text{S}_{1/2}$ & $3\text{p}\ {^{2}}\text{P}_{3/2}$ \\
			\hspace{0.5cm} Mg VIII & 335.2530 & 5.90 & $0.122$ &
				$2\text{s}^{2}\ 2\text{p}\ {^{2}}\text{P}_{1/2}$ &
				$2\text{s}\ 2\text{p}^{2}\ {^{2}}\text{S}_{1/2}$ \B \\
			\hline
			
			\T Fe XVIII & 93.9322 & 6.85 & $1.436\times 10^{-25}, 0.028$ &
				$2\text{s}^{2}\ 2\text{p}^{5}\ {^{2}}\text{P}_{3/2}$ &
				$2\text{s}\ 2\text{p}^{6}\ {^{2}}\text{S}_{1/2}$ \\
			\hspace{0.5cm} Fe VIII & 93.4690 & 5.80 & $0.068$ & 
				$3\text{s}^{2}\ 3\text{p}^{6}\ 3\text{d}\ {^{2}}\text{D}_{3/2}$ & 
				$3\text{s}^{2}\ 3\text{p}^{6}\ 7\text{f}\ {^{2}}\text{F}_{5/2}$ \\
			\hspace{0.5cm} Fe XIV & 93.6145 & 6.30 & $0.177$ &
				$3\text{s}^{2}\ 3\text{d}\ {^{2}}\text{D}_{3/2}$ & 
				$3\text{s}^{2}\ 4\text{p}\ {^{2}}\text{P}_{1/2}$ \\
			\hspace{0.5cm} Fe VIII & 93.6160 & 5.80 & $0.102$ & 
				$3\text{s}^{2}\ 3\text{p}^{6}\ 3\text{d}\ {^{2}}\text{D}_{5/2}$ & 
				$3\text{s}^{2}\ 3\text{p}^{6}\ 7\text{f}\ {^{2}}\text{F}_{7/2}$ \\
			\hspace{0.5cm} Fe XX & 93.7811 & 7.00 & $0.064$ &
				$2\text{s}^{2}\ 2\text{p}^{3}\ {^{2}}\text{D}_{5/2}$ &
				$2\text{s}\ 2\text{p}^{4}\ {^{2}}\text{P}_{3/2}$ \\
			\hspace{0.5cm} Fe X & 94.0120 & 6.05 & $0.292$ & 
				$3\text{s}^{2}\ 3\text{p}^{5}\ {^{2}}\text{P}_{3/2}$ & 
				$3\text{s}^{2}\ 3\text{p}^{4}\ 4\text{s}\ {^{2}}\text{D}_{5/2}$ \B \\
			\hline
		\end{tabular}
	\end{center}	
	\bigskip
	Emission lines identified for analysis in EVE MEGS-A spectra. Each observed emission feature includes the primary line (the first line listed in each section) as well as all other ``secondary'' lines within the full-width at half-maximum of the primary line that have line strengths $>5\%$ of the primary line. Those features with the primary line and wavelength in bold are used to compute DEMs while the other emission features (excluding Fe XVIII) are used for contextual comparison. For the primary line in each emission feature the ``Relative G(T)'' column gives the peak intensity per emission measure (erg cm$^{3}$ sr$^{-1}$ s$^{-1}$) corrected for the elemental abundance (but not weighted by a DEM) as well as the ratio of this value to the Fe IX value. For the ``secondary'' lines only the ratio relative to the associated primary line is given. \label{table:lines}	
\end{table*}

\bibliography{library}%,books,manually_added}
\listofchanges
\end{document}